\documentclass[11pt,a4paper]{article} 
\pdfoutput=1
\usepackage{jheppub}

\usepackage{amsmath, amssymb}
\usepackage{environ}
\usepackage{mathrsfs}
\usepackage{array,arydshln}

\usepackage{graphicx,epsfig}
\usepackage{epic}
\usepackage{color}
\usepackage{youngtab}
\usepackage{float}

\newcommand{\be}{\begin{equation}}
\newcommand{\ee}{\end{equation}}

\newcommand{\mc}{\mathcal }
\newcommand{\Z}{\mathcal{Z}}

\newcommand{\mk}{\mathfrak}

\newcommand{\la}{\longrightarrow}

\def\XXint#1#2#3{{\setbox0=\hbox{$#1{#2#3}{\int}$}
     \vcenter{\hbox{$#2#3$}}\kern-.5\wd0}}

\newcommand{\red}[1]{\textcolor{red}{#1}}

    \newcommand{\beq}{\begin{equation}}
    \newcommand{\eeq}{\end{equation}}
    \newcommand\beqa{\begin{eqnarray}}
    \newcommand\eeqa{\end{eqnarray}}
        \renewcommand{\d}{\partial}

\def \del{ \partial}
\def \la {\label}
\newcommand{\rf}[1]{(\ref{#1})}
\def\ov{\over}
\def\no{\nonumber} 

\def \ci {\cite}
\def \p {\phi}
\def \m {\mu}\def \n {\nu} 
\def \ed {\end{document}}
\def \l {\lambda} \def \r {\rho} 
\def \cP {{\cal P}}
\def \foot {\footnote}
\def \b {\beta} 
\def \dd {{\rm d}} 
\def \om {\omega} 
\def \tr {{\rm tr}}
\def \D {\Delta} 
\def \vp {\varphi} 
\def \OO {{\cal O}}\def \iffa {\iffalse} 
\def \mR {{\mathbb  R}} \def \ha {{{1 \ov 2}}}
\def \bD {{\mathbf{D}}}
\def \wtd {\widetilde}

 \newcommand \ba {\begin{align}}
 \newcommand \ea{ {\end{align} }}

\title{Partition function of free conformal higher spin theory}

\author[a]{Matteo Beccaria} 
\author[b]{, Xavier Bekaert} 
\author[c]{, Arkady A. Tseytlin\footnote{Also at Lebedev Institute, Moscow}}

\affiliation[a]{Dipartimento di Matematica e Fisica Ennio De Giorgi,\\
Universit\`a del Salento \& INFN, Via Arnesano, 73100 Lecce, 
Italy} 

\affiliation[b]{Laboratoire de Math\'ematiques et Physique Th\'eorique\\
Unit\'e Mixte de Recherche $7350$ du CNRS\\
F\'ed\'eration de Recherche $2964$ Denis Poisson\\
Universit\'e Fran\c{c}ois Rabelais, Parc de Grandmont,
37200 Tours, France} 

\affiliation[c]{The Blackett Laboratory, Imperial College, London SW7 2AZ, U.K.}
                     
\emailAdd{matteo.beccaria@le.infn.it}
\emailAdd{xavier.bekaert@lmpt.univ-tours.fr}
\emailAdd{tseytlin@imperial.ac.uk}

\abstract{We compute  the canonical partition function $\Z$  of  non-interacting    conformal  higher spin  (CHS) theory  viewed 
as a   collection of    free   spin $s$  CFT's  in $\mR^d$. 
We discuss in detail the 4-dimensional case (where $s=1$ is  the standard Maxwell vector, $s=2$ is  the Weyl graviton, etc.),
  but also present a generalization  for all even dimensions $d$. 
  $\Z$ may be found   by counting the numbers of conformal operators and their  descendants (modulo  gauge identities and equations of motion)
  weighted by  scaling dimensions. 
 This  conformal operator   counting method requires a careful analysis 
 of  the structure of   characters of   relevant (conserved current,  shadow field and conformal Killing tensor) representations
 of the  conformal algebra $\mk{so}(d,2)$. There is also a close  relation to massless   higher spin partition functions 
  with alternative boundary conditions in AdS$_{d+1}$. 
  The same   partition function $\Z$  may  also be computed from  the CHS  path integral  on a curved   $S^1 \times S^{d-1}$
  background.
 This  allows us  to determine a   simple   factorized  form of the  CHS  kinetic operator  on this conformally flat  background. 
 Summing the individual conformal spin  contributions $\Z_s$ 
  over all spins  we obtain   the total   partition function of the CHS theory. 
  We also find  the  corresponding Casimir  energy on the sphere  and show   that it  vanishes
 if one uses the same   regularization prescription that implies the cancellation  of the  total conformal anomaly $a$-coefficient. 
    This  happens to be true in all  even dimensions $d \geq 2$.
     }

\allowdisplaybreaks

\def  \bes {\be \begin{split} }
\def \ees { \end{split} \ee }

\def \td {\tilde} 
\def \s  {\sigma} \def \cB {{\cal B}} 
\def \nn {{\rm n}}

\def \mso {\mathfrak{so}}

\begin{document}

\begin{flushright}\small{Imperial-TP-AT-2014-04}\end{flushright}

 \maketitle

\flushbottom


\section{Introduction  and summary}

Conformal higher  spin (CHS)  theories  are generalizations  of $d=4$ Maxwell  ($s=1$) 
and Weyl ($s=2$) theories  that describe pure spin $s$ states  off shell, i.e.  have  maximal  gauge  
symmetry  consistent with  locality \ci{Fradkin:1985am} (see also  \ci{Fradkin:1989md,Tseytlin:2002gz,Segal:2002gd}). 
The free CHS  action in flat  4-dimensional space  may be written as 
\be 
S_s= \int d^4x \ \p_s  P_s \, \del^{2s}\,  \p_s  = \int d^4 x\    (-1)^s \, C_{s}  C_{s}  \ , \la{1} \ee
where $\p_s= (\p_{\m_1...\m_s}) \equiv \p_{\m(s)}$ is a totally symmetric tensor and 
$P_s= (P^{\m_1...\m_s}_{\n_1...\n_s})\equiv P^{\m(s)}_{\n(s)}$  is the transverse  projector  which is  traceless  
and  symmetric within  $\mu$ and $\n$ groups of indices.
This  action is thus invariant under a combination of 
differential (analog of reparametrizations)  and algebraic (analog of Weyl)  gauge transformations: 
$\delta \p_s = \del \xi_{s-1}  + g_2 \eta_{s-2}$  (here $g_2$ is  flat   euclidean metric and $\xi$ and $\eta$ are parameter tensors).
  $C_s\equiv C_{\m(s), \n(s)} = (C_{\m_1...\m_s,\n_1 ...\n_s})$  is  the generalized Weyl tensor, {\em i.e.}  the gauge-invariant  field 
strength  that can be written as    
 $C_{\m(s), \n(s)} =  {\cP}^{\l(s),\r(s)}_{\m(s),\n(s)} \del^s_{\l(s)} \p_{\r(s)} $.  Here 
 $\cP_s$ is the  projector\,\foot{Note   the identity 
 $P^{\r(s)}_{\n(s)} \del^{2s} =  {\cP}^{\l(s),\r(s)}_{\m(s),\n(s)} \del^s_{\l(s)}  (\del^s)^{\m(s)}$   
 leading to the second form of the action in \rf{1} \ci{Fradkin:1989md}.} 
  that makes $C_{\m(s), \n(s)} $  totally symmetric and traceless  in each group of indices 
 $\m(s)$ and $\n(s)$   and   also   antisymmetric between them,   so that   $C_{\m(s), \n(s)} $  corresponds to the 
  $(s,s)$  representation of $SO(4)$  described by the   rectangular 
  two-row Young  tableau.\foot{In this paper we always use Young labels $\ell=(\ell_1,\ell_2, ..., \ell_r)$ 
   to denote representation of $SO(2r)$: \   $\ell_i$  are 
  numbers of boxes in rows of  the corresponding  Young tableau.}
   It is  often  convenient    to write  the components of $C_s$ as 
 $C_{\m_1 \n_1  \m_2 \n_2 ... \m_s\n_s}$  with antisymmetry in each pair of  $\m_i$ and $\n_i$   and total symmetry in $\m$'s and $\nu$'s 
  so that   
 $C_1=(F_{\m\n}) $ is the Maxwell tensor, \  $C_2= (C_{\m_1\n_1 \m_2\n_2})$ is the linearized Weyl tensor,  etc. (see 
  \ci{Vasiliev:2009ck,Marnelius:2008er}). 
 
 The analog of  $d=4$  CHS   action \rf{1}  in  any even   dimension   $d$   is
 \be 
S_s= \int d^d x \ \p_s  P_s\,  \del^{2s + d-4} \, \p_s  =   (-1)^s \int d^d x\   \, C_{s}\,  \del^{d-4}\,   C_{s}  \ , \la{2} \ee
 so that  $\p_s$  and $C_s$     have $d$-independent  $SO(d,2)$  scaling dimensions
 \be   \Delta(\p_s) = 2-s \ , \ \ \ \ \ \ \ \ \Delta(C_s)=2    \ . \la{3} \ee
   The action \rf{1},\rf{2}   formally defines a free higher-spin  non-unitary  CFT in  $d$  dimensions. 
   While in this paper we will   discuss only free  CHS theory   which  is a sum of 
   individual   free spin $s$  theories  we shall  emphasize the existence of  its interacting  generalization in the final section.

  Our   aim  here will be to compute the associated  
{\em one-particle} or canonical partition function  $\Z(q)=\sum_{n} \dd_n  q^{\Delta_{n}}$  that counts the numbers of corresponding 
 gauge-invariant  conformal primaries and their descendants  weighted with scaling dimensions 
 like in the familiar $d=4$  free standard  scalar   and spin 1 cases in \ci{Cardy:1991kr,Kutasov:2000td}.\foot{Note  the the $s=0$ field 
 which is the member of the  CHS family is  non-dynamical  in $d=4$   and is 
  the same as the standard 2-derivative conformal scalar field only in $d=6$.}

 One   should  also  find  the same $\Z_s(q)$   from   the standard  finite  temperature  one-loop  partition function $Z_s$ on  
  $S^1  \times S^{d-1}$   background (with euclidean time  circle of length $\b$)  
 which may be interpreted as {\em multi-particle} or    grand  canonical partition function given  by 
 \be 
 \ln Z_s = \sum_{m=1}^\infty { 1 \ov m} \Z_s (q^m) \ , \ \ \ \ \ \ \ \ \qquad    q= e^{-\b} \ . \la{4} \ee
 To compute $Z_s$  then  requires   the knowledge  of the kinetic  CHS operator on  curved  $S^1 \times S^{d-1}$   background. 
 
The form  of the  covariant   kinetic  CHS  operator  $\OO_s= D^{2s+d-4}  + ... $ on a 
 curved background   is not known in general\,\foot{This operator   is expected to     be reparametrization and Weyl invariant 
   and consistent with  CHS   gauge symmetries for any    background 
 metric  solving Bach  equations of Weyl gravity theory.}
but  it should   have  a particularly simple  structure  on a conformally flat  space. Indeed, it was  found recently 
  that  on a  conformally flat    background   which is also an Einstein space, like $S^d$ or AdS$_d$, the operator  
$\OO_s$  factorizes  into  a product of  2-nd derivative  partially massless  and massive spin $s$ operators    \ci{Tseytlin:2013jya,
Tseytlin:2013fca,Metsaev:2014iwa,Nutma:2014pua}. 

Below  we will   determine  the form  of $\OO_s$ on the conformally flat 
but  non-Einstein  $d=4$   background  $S^1  \times S^{3}$   by  first   (i)  
finding it   explicitly  in  the familiar  $s=1$ and $s=2$ cases,  then     (ii)    conjecturing a 
 natural generalization to the $s >2$ case,   and finally 
(iii)  checking the consistency of the resulting partition function $\Z_s$ with   the  one   found by  direct 
count  of  conformal CHS operators in  $\mR^4$ 
 that can be   justified  by representation-theoretic  methods.
 We will also  find  the expression for  $\Z_s$  for all even dimensions  $d>4$.

The  study of  this     partition function 
 is also of interest in  the context  of remarkable relations   between   conformal higher spin theory in $d$  dimensions, 
singlet sector of  free   scalar CFT in  $\mR^d$   and dual  massless higher spin   theory in AdS$_{d+1}$. 
 A free massless   complex scalar theory  $ S= -\int d^d x \,   \Phi^*_r \del^2  \Phi_r $  \ ($r=1, ..., N$)
has  a tower of (on-shell) conserved  symmetric traceless  higher spin currents  
 $J_s \sim  P_s \Phi^*_r  \del^{s}   \Phi_r, \   \del J_s=0$  which are conformal fields of dimension $\Delta(J_s)= s + d -2\equiv \Delta_+ $. 
 Adding these  currents to  the action with the  {\em source}  or  {\em shadow}  fields $\p_s(x)$  
  one observes that this  $\p_s$ has the 
 same   dimension   $2-s\equiv \Delta_-= d -\Delta_+$ and  effectively  the same algebraic and gauge  (due to   properties of $J_s$)
 symmetries   as  the  CHS field in \rf{3}.
   Integrating  out the  free fields $\Phi_r$ in the path integral  then gives 
 an  effective action for $\p_s$ the 
 leading local (logarithmically divergent)  part of which is, 
at quadratic level, 
 the same as the  
 classical  CHS action in \rf{1},\rf{2} \ci{Liu:1998bu,Tseytlin:2002gz}.
 From the AdS/CFT perspective this induced action  should  be  found upon the  substitution 
 of the solution of the  Dirichlet problem (with $\p_s$ as the boundary data) 
 into   the  classical action of a massless spin $s$ field in AdS$_5$.

 In addition to this  {\em classical}  relation, there is also a {\em one-loop} one 
  \ci{Giombi:2013yva,Tseytlin:2013jya} 
 \be 
 { Z_{-\,  s}  \ov Z_{+\,  s} }  \Big|_{M^d}  =    Z_s  \Big|_{M^d}   \  . \la{5}
  \ee
 Here $Z_s$   is the 1-loop CHS  partition function  on a conformally flat  space $M^d$. 
 $ Z_{+\, s}$  is  the free   scalar CFT   partition function  in the spin $s$ current  part of the  singlet   sector
 (the total singlet sector partition function is $\prod_s Z_{+\, s}$)  and  
 $ Z_{-\, s}$   is  its spin $s$  shadow operator   counterpart.\foot{This relation  can be  motivated  \ci{Giombi:2013yva}
 by considering the  {\em double-trace}   $J_s J_s$ deformation of  the  free  large $N$   scalar  theory
 under which the scaling  dimension of only one  (spin  $s$) operator  is changed, i.e. the l.h.s.  of \rf{5} is 
 $Z_{\rm UV}/Z_{\rm IR}$, i.e. 
 the ratio  of the UV and IR large $N$   fixed point CFT partition functions (this argument can be made precise in $d=3$
 \ci{Giombi:2013yva}). 
 In even $d$  case that we are considering here $Z_s$ is  given   by the  determinant  of the local CHS  kinetic operator while 
 in odd $d$   the corresponding kinetic operator is the full nonlocal  kernel $\langle J_s  J_s \rangle$ appearing in the
  induced theory.} 

 By  the  AdS/CFT rules,   $Z_{\pm \,  s}$  should  have   the 
  equivalent  interpretation   as the one-loop 
 partition function  $Z^{(\pm)}_{  s}$
 of the  massless higher  spin $s$ field  $\vp_s$ in AdS$_{d+1}$  with boundary $M^d$  
 computed with the standard
  $\vp_s \sim z^{\D_+ -s} $ or the alternative  $\vp_s \sim z^{\D_- -s} $ boundary conditions (see  \ci{Giombi:2013yva}   
   and references there). 
   Thus,  we should also  have  the following relation between the  massless  higher spin partition functions 
   in the bulk AdS$_{d+1}$ and the  CHS   partition function  at the  conformally flat boundary $M^d$
   \be 
 { Z^{(-)}_s  \ov  Z^{(+)}_s  }  \Big|_{  AdS_{d+1}  }   =    Z_s  \Big|_{M^d}   \  . \la{6}
  \ee
   These  relations  \rf{5},\rf{6}   were    verified   explicitly  \ci{Giombi:2013yva,Tseytlin:2013jya,Tseytlin:2013fca,Giombi:2014iua} 
     in the case of $M^d$  being the sphere $ S^d$    for even 
   $d=4$ and $d=6$  where  $\log Z_s$ is essentially determined by the   conformal anomaly $a$-coefficient.  
   
   Below we will   demonstrate  the  validity of  these relations  also in the case of $M^d = S^1 \times S^{d-1}$. 
   In this case  Eqs.~\rf{5}  or \rf{6}   may be written as  a   relation   between  the  corresponding 
    one-particle partition functions as functions of $q= e^{-\b}$
    (cf. \rf{4})
   \be 
   \Z_{-\,  s} (q) -  \Z_{+\,  s} (q)   =    \Z_s (q) \ , \ \ \ \ \ \ \ \ \  \qquad   \Z_{\pm \,  s}(q) =  \Z^{(\pm)}_s (q)
     \ . \la{7} \ee
   The expression   for $ \Z_{+\,  s} $  is  straightforward  to find  using the conformal operator counting method  
   in $\mR^d$ \ci{Cardy:1991kr,Kutasov:2000td}: 
    it should compute  the  operators  represented   by the  components 
   of the traceless   symmetric   spin $s$ current operator $J_s= (J_{\m_1 ...\m_s})$  of dimension 
   $\D_+$   and all of  its conformal descendants (derivatives)    modulo   the  conservation  condition $\del^{\m_1} J_{\m_1...\m_s} =0$ 
   (rank $s-1$ tensor of dimension $\D_+'= \D_+ +1$) and all of its derivatives. This immediately leads to 
   \ba
   & \Z_{+\,  s}  =  { \nn_s\, q^{\D_+}  - \nn_{s-1}\,   q^{\D'_+}   \ov (1-q)^d } \ , \ \ \ \ \ \ \ \la{8}\\  
     \D_+ =  s + d-2 \ , \ \ \ \ &  \D'_+  =  s + d -1  \ ,   \ \ \ \ \  
       \nn_s = ( 2 s + d-2)\,   { (s + d-3)! \ov (d-2)! \, s!}  \la{9} \ , 
   \end{align}
   where $\nn_s$  is the number of components of totally symmetric traceless rank $s$ tensor in $d$ dimensions. 
   Explicitly, in $d=4$  this gives   
   \be
   d=4: \ \ \ \ \ \qquad    \Z_{+\,  s}  =  { (s+1)^2\, q^{s+2 }  -   s^2\,   q^{s+3}   \ov (1-q)^4 } \ , \ \ \ \ \ \ \ \la{88}\ee
  Eq.~\rf{8}  is   indeed the same  as the massless   spin $s$ field  partition function    $\Z^{(+)}_s   $ 
   in   thermal  AdS$_{d+1}$  with $S^1 \times S^{d-1}$  boundary   \ci{Gopakumar:2011qs,Gupta:2012he,Giombi:2014yra},   
    with the negative (current conservation  subtraction)   term here corresponding to the ghost contribution.\foot{Eq.\rf{8} has 
   also the   interpretation of the character of  the short representation of the $SO(d,2)$ conformal group  with 
    dimension $\D_+$ and spin $s$, i.e.  $\chi_{(\D_+,s,0,...,0)} (q, 1,...,1)$  which is also 
    a  difference of characters of the  corresponding 
   long representations $(\D_+,s,0,...,0)$ and $ (\D_+ + 1,s-1,0,...,0)$  (see    \ci{Dolan:2005wy}  and appendix \ref{A:BGG} below).}
  
  A natural   guess 
      for the expression   for $ \Z_{-\,  s} $  that should   be counting  
   the shadow  spin $s$ operators (modulo gauge degeneracy) is just to replace the dimensions $\D_+$  and $\D_+'$ in \rf{8} 
   by their {\em shadow} ($\Delta\to  d -\Delta$)   values, {\em i.e.} 
   \be
   \wtd  \Z_{-\,  s}  =  { \nn_s\, q^{\D_-}  - \nn_{s-1} \,  q^{\D'_-}   \ov (1-q)^d } \ , \ \ \ \ \ \ \  \ \ \ \ \ \  \D_- = 2 -s\ , \ \ \ \ \ \D_-'= 1-s \ .  
   \la{10}\ee
As we shall discuss below, this $\D_+ \to \D_-$   prescription that  was used  in the 
$S^d$ case \ci{Giombi:2013yva}   here  does not give the full  answer: 
the  expression in \rf{10}  satisfying $ \widetilde  \Z_{-\,  s}(q) = (-1)^d  \Z_{+\,  s}(1/q)$  
  is to be corrected   by an extra contribution $\sigma_s(q)$ that removes, in particular, all negative powers of $q$
 from the small $q$ expansion. This extra term  
  may be interpreted as the character of the finite dimensional irreducible representation of $SO(d,2)$ 
  corresponding to the conformal Killing tensors in $d$ dimensions. This is related  to 
   gauge degeneracy associated 
  with shadow fields. 
 We shall also provide a general group-theoretic argument for counting  of shadow  operators based on 
 characters of relevant  conformal  algebra   representations  using the  general  analysis in 
\ci{Shaynkman:2004vu}.
\foot{For other  discussions of shadow fields  in the context of AdS/CFT see also 
 \ci{Metsaev:2008fs,Metsaev:2009ym,Metsaev:2011uy,Bekaert:2012vt}.}
 
  Explicitly,    we will find  that  in $d=4$
 \be 
    \Z_{-\,  s}  =  \td  \Z_{-\,  s} (q)   + \s_s (q) 
     = { 2(2\,s+1)\,q^{2}-(s+1)^{2}\,q^{s+2}+s^{2}\,q^{s+3}  \ov (1-q)^4}  \  .  \la{11} 
\ee        
 Then,  from \rf{88} and \rf{11},  
 \be 
      \Z_{-\,  s}  - \Z_{+\, s} = {
2(2\,s+1)\,q^{2}-2 (s+1)^{2}\,q^{s+2}+2 s^{2}\,q^{s+3}  \ov (1-q)^4}
  \ . \la{12} \ee
 This turns out to be    the same as the CHS   partition function in $d=4$, or,  equivalently, 
\be d=4: \ \ \ \ \ \qquad   \qquad \ \ 
\Z_s =         { 2 q^2 \big[ 
 (s+1)^{2} ( 1 - q^{s})-  s^{2} (1- q ^{s+1} )    \big]     \ov (1-q)^4}           \ , \qquad   \la{13} 
\ee
 thus verifying the relations \rf{5},\rf{7}. 
 
 Moreover, we  will see  that   one can  give  a natural 
    interpretation  to  $\Z_{-\,  s} $   and $  \Z_{+\, s}$ in terms of counting of 
 conformal   operators  in  the CHS theory \rf{2}   in $\mR^d$:   
 \be 
 \Z_{-\,  s}  = \Z^{\rm off-shell}_s \ , \ \ \ \ \ \ \    \ \ \    \Z_{+\,  s}  = \Z^{\rm e.o.m.}_s  \ , \ \ \ \ \ \ \ \ \ \ 
 \Z_s = \Z^{\rm off-shell}_s -   \Z^{\rm e.o.m.}_{s} \ .  \la{14} \ee 
 Namely, the shadow partition function 
 $\Z_{-\,  s} $    counts (derivatives of)  off-shell components of 
  field strength   $C_s$  (and  its  derivatives)  modulo non-trivial gauge identities 
  while  
  $\Z_{+\,  s} $  counts the components  of  the  (derivatives of) equations of motion   for $C_s$  following from 
  the  CHS action \rf{2} upon variation   over $\p_s$   (also modulo   identities)
  that should be subtracted to get  the physical  {\em on-shell}  result for $Z_s$. 
  The  {\em kinematical}  or  {\em off-shell}  relation between $\Z_{-\,  s} $   and $\Z_s$  may not be    unexpected as the shadow 
  spin $s$ operators   and the  conformal higher spins  have the same  symmetries and dimensions. 
  
  For example, in $d=4$  each of the terms in the numerators of \rf{11},\rf{88} and \rf{12}  has a clear meaning: 
   (i) $2(2\,s+1)q^2$ represents  the  components of  the   CHS field   strength $C_s$  of dimension 2
    ($6 q^2$ for $s=1$  Maxwell field);
   (ii)  $(s+1)^{2}q^{s+2} $  corresponds to the  components  of the  equations of motion tensor $ B_s = \del^s C_s $ 
   and also of the gauge identity tensor $\cB_s = (\epsilon_4)^s  \del^s C_s $ which  are both 
   symmetric traceless rank $s$ tensors  of dimension  $s+2$ 
   (generalizations  of $ \del^\m F_{\m\n}$ and $\epsilon^{\m\n\l\r} \del_\n F_{\l\r}$ for $s=1$);
   (iii)  $ s^{2}\,q^{s+3}$ term accounts for   the trivial identies $\del B_s =0$ and $\del \cB_s=0 $ (to be added back to avoid overcounting) 
   which are  symmetric traceless rank $s-1$ tensors   of dimension $s+3$. 
   
   The  triality relations between  different objects and partition functions discussed above are illustrated by  the diagram below. 
\begin{center}
\vskip -30pt
\begin{picture}(200,200)
\put(100,100){\makebox(0,0){$(\Z_{+\, s}, \Z_{-\,s})$}}
\put(120,120){\vector(1,1){35}} 
\put(130,160){$(\partial^{s}C_{s}, C_{s})$}
\put(80,120){\vector(-1,1){35}}
\put(-10,160){$({\rm current}\,  J_s,\,  {\rm shadow}\,  \td J_s)$}
\put(100,80){\vector(0,-1){40}}
\put(50,20){$(\log \det_{(+)}, \log \det_{(-)})$}
\put(-30,125){free scalar ${\rm CFT}_{4}$}
\put(145,125){conformal higher spin $s$\  ${\rm CFT}_{4}$}
\put(110,60){massless  higher spin $s$ in AdS$_5$}
\end{picture}
\end{center}
\vskip -20pt
   Finally,  we may  obtain the total CHS partition function  by summing over all spins. 
   Assuming a natural regularization discussed  in section 6  we   find   that 
     \be 
 d=4: \ \ \ \   \Z(q) =\sum_{s=0}^\infty \Z_s(q) =  -\frac{q^2 \left(11 +26 q+11q^2\right)}{6 (1-q)^6}  \  , \ \ \ \ \ \ \   \Z(q) = \Z(1/q) \ . \la{1100}\ee
 This implies also   the   vanishing  of the associated  total  Casimir energy on $S^3$, as 
  in the case  of the massless   higher spin partition function 
 $ \Z_+(q) =\sum_{s=0}^\infty \Z_{+\, s}(q) =  \frac{q^2 (1+q)^2}{(1-q)^6} $ discussed in \ci{Giombi:2014yra}. 
   
   \medskip
   
This paper is organized as follows. 
We shall   start in section 2   with a review of the  two equivalent  methods of computing   the canonical partition  function of a free CFT: 
the conformal operator counting  method  in $\mR^d$    \cite{Cardy:1991kr,Kutasov:2000td}
and   the  finite temperature computation  on  a  spatial sphere $S^{d-1}$. We will illustrate these methods on the examples of 
standard conformal   scalar in $d$ dimensions and Maxwell theory in $d=4$. 

In section   3, we shall generalize the discussion of the spin  1   case  to 
 the    conformal spin 2 case in $d=4$,  i.e.  Weyl gravity theory. 
 We shall   first compute  the   corresponding  one-loop partition function $\Z_2$  on $S^1 \times S^3$  by 
expanding the non-linear Weyl  action (both in 4-derivative and  an equivalent 2-derivative formulations) 
 to quadratic order in fluctuations  and  
  then check the agreement of the result with the one found   by the  operator counting method 
for the  Weyl  action  linearized   near flat space. 

In section 4 we   shall  use the  above explicit $s=1,2$  results  as a motivation  for a 
   a proposal   for the factorized  form  of the conformal higher spin $2s$-derivative 
operator on the $S^1 \times S^3$    background   and for   the structure of the associated partition function.
We  will  then  support  the  consistency of our proposal by   demonstrating 
    the agreement of the  resulting canonical  partition function $\Z_s$   with 
the one that can be  found directly  by   counting  gauge-invariant conformal operators   corresponding to 
 the  linearized conformal  higher spin action \rf{1} in $\mR^4$.

Next, in section 5.1,   we shall  clarify the 
relations \rf{7},\rf{14}  between 
  the  conformal  spin $s$ partition function  and the  partition functions 
associated with   spin $s$  conformal  operators  in  the singlet sector of free 
  CFT in $d=4$ and  thus  also with  the one-loop partition  functions  of massless    spin $s$ field in AdS$_5$.
We will  suggest a method to compute the shadow field  partition function $\Z_{-\, s}$  \rf{11} 
 that will allow a straightfoward 
generalization to $d>4$ (section 5.2) 
and  will   thus lead to the expression for  the conformal higher spin partition function  $\Z_s$ 
 generalizing \rf{13} to any even  dimension    $d $. 
In  section 5.3,    we shall  explain how our general result \rf{512} for  the conformal higher spin partition function 
 can be derived in a rigorous way  in terms of  characters  of the relevant   Verma modules of the conformal algebra $\mso(d,2)$ 
 (with details presented in  appendix \ref{A:BGG}). We shall also comment on the special case of $d=2$.

 Finally, in section 6,   we shall  sum the individual  contributions $\Z_s$  to obtain the total  (summed over all spins) 
 partition function of 
 the conformal higher spin theory.  We shall also find the   Casimir  energy on $S^{d-1}$   and show   that it vanishes
 if one uses the same   regularization prescription that implies the vanishing of the  total conformal anomaly $a$-coefficient 
  \ci{Giombi:2013yva,Tseytlin:2013jya,Tseytlin:2013fca,Giombi:2014iua}.  This  happens to be true in any even dimension $d \geq 2$. 
 
 Section 7 contains some concluding remarks, empasizing  the existence of the full non-linear generalization of the 
 conformal higher spin action \rf{1},\rf{2}  viewed as an induced action of a free CFT$_d$ in  background fields   dual to 
 all conserved spin $s$ currents. 
 
 There are also several technical appendices. In particular, appendix F  (using  the results of \cite{Shaynkman:2004vu})
   explains  the structure  of spaces  of  representations   and associated characters 
   of the conformal  algebra  in $d$ dimensions  that  are relevant for the 
 computation of the  CFT partition functions in the main text.



\section{Free CFT partition function:    scalar and vector  examples} 

We shall   start  with a review of the  two equivalent  methods of computing   the canonical partition  function of a free CFT: 
the operator counting  method  in $\mR^d$    \cite{Cardy:1991kr,Kutasov:2000td}
and   the  finite temperature computation  on  a  spatial sphere $S^{d-1}$. 

Radial quantization   relates    conformal operators in $\mR^d$     with dimensions $\Delta_{n}$ 
to eigenstates  of the Hamiltonian  (dilatation operator)
on $\mR_t \times S^{d-1}$.
Given the spectrum  of    eigenvalues
 $\om_n= \Delta_{n}$  and their degeneracies $\dd_n$,    
   the  ``one-particle'' or canonical  partition function  is 
\be
\Z(q) = \tr\, e^{-\b H} =  \sum_n  \dd_n\, e^{-\b \om_n} =   \sum_n  \dd_n\,   q^{\D_n}   \ , \ \ \ \ \ \ \ \ \
q\equiv e^{-\b} 
  \ .  \la{21} \ee
The    multi-particle or grand canonical  partition function    is then   given by 
\be  \ln Z  = - \sum_n \dd_n \ln  (  1 - e^{-\b \om_n})  = \sum_{m=1}^\infty  { 1 \ov m}  \Z(q^m)  \ . \la{22}  \ee
At the same   time, \rf{22}   can be  found  also from the     free QFT path 
integral on $S^1_\b \times   S^{d-1}$,  {\em i.e.}   computing  the determinant   of the corresponding Laplacian $\OO$.
We shall review the  computation of $\Z$   below    on the examples  of conformal scalar  in $d$ dimensions and  Maxwell vector in $d=4$.
The latter is the first  non-trivial representative  of the   conformal higher spin tower  in 4 dimensions.


\subsection{Operator counting method in $\mathbb R^d$}

\iffa 
Thermodynamics of a CFT on $S^{3}$ is governed by the canonical partition function 
\be
\Z = \mbox{Tr} e^{-\beta\,H},
\ee
where $\beta$ is the inverse temperature, and $H$ the Hamiltonian. We can conformally 
map $\mathbb R\times S^{3}$ to $\mathbb R^{4}$ and the Hamiltonian is mapped to the generator of scaling transformations in flat  space. Compactifying one dimension on a circumference of radius $\beta$, 
$Z$ can be identified with the generating functional of conformal dimensions on $\mathbb R^{4}$
\be
\label{KutLarSum}
\Z = \sum_{\Delta}q^{\Delta},\qquad q=e^{-\beta},
\ee
where we take unit radius in $S^{3}$.
 Hence, one can evaluate $\Z$ by performing the quadratic path integral 
associated with the wave equation on $S^{1}\times S^{3}$, or by counting local operators in flat  space CFT.
In the simple case of a conformally coupled scalar, counting is very simple. 
\fi 

Let us start with a    free   massless  scalar field  $\Phi $  in $\mR^d$ with   the standard action $\int d^d x \, (\del \Phi)^2$ 
 and thus  dimension $ \Delta(\Phi)= \ha (d-2)$. The  
 lowest  dimension   conformal  operator  is the scalar  field  itself   contributing $q^{ \ha (d-2)}$ to the sum in \rf{21}. 
Its conformal descendants  are found by adding   derivatives:
$\del_{\m_1} ...\del_{\m_k}\Phi$.   As derivatives in all $d$ dimensions are independent 
 and each power of derivative in a given direction enters only once   we get a   factor 
 $\sum_{k=0}^\infty q^k  = (1-q)^{-1}$ from each of the $d$ directions. 
This   counting ignores the fact   that some operators vanish due to equations of motion $\del^2 \Phi=0$. 
Since $\Delta( \del^2 \Phi) = \ha (d-2)   + 2$   we thus need to subtract a term $q^{ \ha (d-2)+2 }$, 
 {\em dressed}  again  by extra derivative factor $(1-q)^{-d}$. 
%
%
The resulting  partition function  of a conformal scalar is then 
\be  \Z_{\rm c.s.}(q) = \frac{q^{ d-2 \ov 2  } ( 1 - q^2)}{(1-q)^{d}} \ , \ \ \ \ \ \ \ \ \ \ \ 
\Z_{\rm c.s.}(q)\Big|_{ d=4}= \frac{q-q^{3}}{(1-q)^{4}}\ .   \la{23}
\ee
This  one-particle partition function  corresponds  to the character of the free scalar (Dirac singleton) 
representation of the conformal group $SO(d,2)$  (see  \cite{Dolan:2005wy}). 

Next, let   us consider  the standard Maxwell  vector field  in $d=4$. 
Here  lowest  dimension  gauge-invariant operator  is the  field strength $F_{\mu\nu}$ with dimension 
$\Delta=2$ and 6 components,  giving  a term $6q^{2}$.  Its derivatives  give extra factor of $(1-q)^{-4}$. 
This overcounts by ignoring the   vanishing of some operators due to  equations  of motion $\partial^{\mu}F_{\mu\nu}=0$
and gauge  identities $\partial^{\mu}F^{*}_{ \mu\nu}=0$  (and their derivatives). This suggests  
subtraction of $ -(4+4)q^{3}$ times $(1-q)^{-4}$, but this also  overcounts as some identites are trivial, descending 
from the two   $\partial_{\mu}\partial_{\nu}F^{\mu\nu}=0$   and  $\partial^{\mu}\partial^{\nu}F^{*}_{\mu\nu}=0$
corresponding to operators of dimension 4. 
This  requires 
adding  back 
$2q^{4} (1-q)^{-4}$.
The final  $d=4$   vector partition function is then 
\be
\Z_{1}(q) = \frac{6q^{2}- 8 q^{3}+2q^{4}}{(1-q)^{4}} = \frac{2\,(3-q)\,q^{2}}{(1-q)^{3}} \ . \la{24}
\ee
This counting argument can be straightfowardly generalized  to the case of a conformal  vector 
in even $d$ dimensions  with the  action \rf{2}, i.e. $S_1= - { 1 \ov 4} \int d^d x \    F_{\m\n} \del^{d-4}  F_{\m\n}$
(see section 6 below).  

\

\subsection{Partition function  on  $S^1\times S^{d-1}$} 
\label{sec:spin0}
Let us now review how  the same expressions for $\Z$  can be found by computing the standard QFT partition function on a curved 
$S^1_\b \times S^{d-1}$  background  (we assume that $S^{d-1}$ has unit radius). 
 The partition function of a free conformally coupled
scalar  is 
\be\la{25} 
- \log Z_{\rm c.s.}  = \frac{1}{2}\,\log\,\det\, \OO_0  \ , 
\qquad \qquad \OO_0  = -D^{2}+\frac{d-2}{4\,(d-1)}\,R \ . 
\ee
On  $S^{1}\times S^{d-1}$ 
we have   
\be \la{255}
D^{2} \equiv  D_{\m}D^{\m}= \del_0^2  + \bD^2 \ , \ \ \ \ \ \ \ \     
\bD^2= \bD^i\bD_i= D^2_{S^{d-1}}   \ , \ee
where   $\del_0$   is derivative along compact euclidean time direction of length $\b$. 
The  scalar curvature  is $R= R(S^{d-1})  = (d-1)(d-2)$, so  that 
\be
\OO_{0} = -\partial_{0}^{2}-\mathbf{D}^{2}+\frac{1}{4}\,(d-1)^{2}\ .   \la{26}
\ee
The eigenvalues  of the Laplacian $-\mathbf{D}^{2}$  on $S^{d-1}$ 
and their multiplicities are (see appendix  \ref{A:Laplace}) 
\begin{align}
\lambda_{n}(S^{d-1}) =& n\,(n+d-2), \la{27} \\
\dd_{n}(S^{d-1})  
=&(2n+d-2)\,\frac{(n+d-3)!}{n!\,(d-2)!}.\la{28} 
\end{align}
Hence, the eigenvalues of \rf{26} are 
\be
\lambda_{k,n} = w^2 +\omega_{n}^{2}, \qquad   w= \frac{2\pi k}{\beta} \ , \quad 
\omega_{n} = n+\frac{1}{2}(d-2), \qquad  k\in \mathbb Z, \quad n = 0, 1, 2, \dots.\la{29} 
\ee
Then, computing $-\ln Z_{\rm c.s.}= \ha \log \det \OO_0=   \ha \sum_{k,n} \dd_n \log \lambda_{k,n} $  one finds  by the standard argument 
the expression in \rf{22}   where 
\be
\Z_{{\rm c.s.}}(\beta) = \sum_{n=0}^{\infty}d_{n}\,e^{-\beta\,\big[ n+\frac{1}{2}(d-2)\big]} = \frac{q^{\frac{d-2}{2}}(1-q^{2})}{(1-q)^{d}}\ ,  
\la{210} \ee
which is indeed  the same  as in \rf{23}.

The  quantization of the  $d=4$   Maxwell   action $S_1= -\frac{1}{4}\ \int d^4 x \sqrt g \,  F_{\mu\nu}\,F^{\mu\nu}$
   in curved   background 
in covariant Lorentz gauge  gives the following  well-known expression  for the vector field   
partition function 
\be
Z_1 = \frac{\det(-D^{2})}{\left[\det(- g_{\mu\nu} D^{2}+R_{\mu\nu})\right]^{1/2}}\ , \la{211} 
\ee
Specializing to $S^1 \times S^3$  where $R_{00}=0, \ \  R_{ij} = {1 \ov 3} g_{ij} R, \ R=6$, and  
    $A_{\mu} = (A_{0}, A_{i})$  \ ($i,j=1,2,3$) 
we get from \rf{211} 
\be\la{212}
Z_1 = \left[\frac{\det(-D^{2})}{\det(-g_{ij} D^{2}+ R_{ij})}\right]^{1/2}.
\ee
Splitting the 3-vector field operator  into the  transverse ($D^{i}A_{i,\perp}=0$)  and longitudinal parts we end up with 
(using that $R_{ij} = 2\,g_{ij}$)
\be
\label{213}
Z_1 = \frac{1}{\big[\det(-g_{ij} D^{2}+  R_{ij} )_{\perp}\big]^{1/2}}
=  \frac{1}{\big[\det \, \OO_{1\, \perp}\big]^{1/2}} \ , \ \ \ \ \ \ \ \ 
\OO_{1\, ij} = (- \del_0^2  - \bD^2 + 2 )_{ij} \ , 
\ee
where $\OO_{1\, \perp}$   is defined on transverse  3-vectors. 

The same   expression   can be obtained  directly 
by choosing the temporal gauge $A_{0}=0$  in the  original path integral. In $S^1 \times S^3$  case 
the corresponding ghost  factor is $\det(\partial_{0})$
while the  Lagrangian is   $\mathscr L = -\frac{1}{4} F_{\mu\nu}\,F^{\mu\nu}
= -\frac{1}{2} \partial_{0}A_{i}\,\partial_{0}A^{i}-\frac{1}{4} F_{ij}\,F^{ij}
$.
Changing variables  $A_{i} = A_{i\,  \perp}+D_{i}\,\vp$ introduces the Jacobian  factor 
$\big[\det(-\mathbf{D}^{2}) \big]^{1/2}$ 
while   the Lagrangian  becomes (up to a total derivative) 
\be
\begin{split}
\mathscr L 
= -\frac{1}{2} \varphi\,\partial_{0}^{2}\mathbf{D}^{2}\,\varphi
-\frac{1}{2} A^i_{\perp}(- g_{ij}D^{2}+R_{ij})
\,A^{j}_{\perp}. \la{215} 
\end{split}
\ee
Integration over $\varphi$ gives the  contribution 
$[\det(-\partial_{0}^{2}\,\mathbf D^{2})]^{-1/2}$
that 
cancels  the product of the ghost   and Jacobian factors.  
The final result
is thus   again \rf{213}.

Using   the eigenvalues  and  their multiplicities  
of the transverse vector  Laplacian $(-\bD^{2})_{1\,  \perp}$   on $S^3$   given by \rf{a1}--\rf{a4},
we conclude  that the spectrum   of $\OO_{1\, \perp}$   in \rf{213}   is ($\d_0\to i w, \ w= { 2 \pi k \ov \b}$, \ cf. \rf{29})
\be
\lambda_{k,n} =w^2 +  (n^{2}+4n+2)+2 = w^2 + \omega_n^2 \ , \quad 
\  \omega_n= n+2\ ,\qquad  \dd_{n} = 2(n+1)(n+3) \ , \la{216}
\ee
and thus   the  one-particle partition function  corresponding to $Z_1$ in \rf{213}   is  given by 
\be
\Z_{1}(\beta) = \sum_{n=0}^{\infty} \dd_{n}\ e^{-\beta(n+2)} = \frac{2\,(3-q)\,q^{2}}{(1-q)^{3}} \ .  \la{217}
\ee
This  is  again  in agreement with the  expression  \rf{24}  found  by   the  operator counting method.

\section{Conformal spin 2  in $d=4$}
\label{sec:spin2}

Let us  now   consider  the    conformal spin 2 case in $d=4$,  {\em i.e.}  Weyl gravity    with the 
full non-linear action  being  (we drop total derivative)
\be
\label{31}
S_2 =\ha  \int d^{4}x \,\sqrt{g}\,C_{\m\n\l\r} C^{\m\n\l\r} = \int d^{4}x\,\sqrt{g}\,\Big(R_{\mu\nu}R_{\mu\nu}-\frac{1}{3}\,R^{2}\Big) \ . 
\ee
Here,  we shall   first compute  the   corresponding  one-loop partition function on $S^1 \times S^3$  by 
expanding the action \rf{31} to quadratic order in fluctuations near this conformally flat  background. We shall  
  then check the agreement of the result with the one found   by the  operator counting method 
for the linearized   action \rf{31} expanded near  $\mR^4$.


\subsection{Quadratic fluctuation operator in  conformally flat  background}

Since we  are   interested  in  quantizing spin 2 fluctuations on the  
conformally flat   $S^1 \times S^3$  background\,\foot{This  background  solves the Bach equations
 of motion corresponding to 
\rf{31} so that the resulting partition function will be gauge-independent.},
  in expanding \rf{31} we may  ignore terms with the 
  Weyl tensor of the background metric. 
We may  also drop  terms with  covariant derivatives of the curvature. 
Then,  using  the expressions in  appendix \ref{A:expan}  and  assuming the reparametrization and Weyl  gauge conditions 
$D_{\mu}h^{\mu\nu}=0$, \ $h_{\mu}^{\ \mu}=0$, 
we find from \rf{31}  the following quadratic fluctuation Lagrangian
\be
\begin{split}
\label{33}
\mathscr L^{(2)} &=\frac{1}{4} D^{2}h_{\mu\nu}\,D^{2}h^{\mu\nu} -R^{\mu}_{\ \rho} h_{\mu\nu}\,D^{2}h^{\nu\rho}\,+\frac{1}{2}\,R^{\mu\nu}\,
h_{\alpha\beta}\,D_{\mu}D_{\nu}\,h^{\alpha\beta} \\
 &-\frac{3}{2}\,  R_{\rho\sigma}\, R^{\sigma\mu} \, h_{\mu\nu}\,h^{\nu\rho}\,
 + \frac{1}{2}\,R^{\nu\rho} R^{\sigma\mu}  h_{\mu\nu}\, \,h_{\rho\sigma}\, 
 +\frac{1}{6}\,(h_{\mu\nu}\, R^{\mu\nu})^{2}
+\frac{1}{4}\,R_{\mu\nu}\, R^{\mu\nu}\,h_{\alpha\beta}\,h^{\alpha\beta}\\
&+\frac{1}{2}\,R\, R_{\rho}^{\ \mu} \,h_{\mu\nu}\,h^{\nu\rho}\, -\frac{1}{9}\,R^{2}\,h_{\mu\nu}\,h^{\mu\nu}.
\end{split}
\ee
 In the special case 
  when our conformally flat  background is also an Einstein space  $R_{\mu\nu}=\frac{1}{4}R\,g_{\mu\nu}$, {\em i.e.} 
  for $S^4$ or AdS$_4$, 
the Lagrangian \rf{33}  reduces to 
\be\la{34} 
\mathscr L^{(2)}_{R_{\mu\nu}=\frac{1}{4}Rg_{\mu\nu} } = \frac{1}{4}D^{2}h_{\mu\nu}\,D^{2}h^{\mu\nu}
-\frac{1}{8}\,R\,h_{\mu\nu}\,D^{2}
h^{\mu\nu}+\frac{1}{72}\,R^{2}\,h_{\mu\nu}\,h^{\mu\nu} = \frac{1}{4}\,h^{\mu\nu}\,\td \OO_2\,h_{\mu\nu},
\ee
where  the 4-th order operator  $\mc O$   defined on transverse traceless  tensors $h_{\m\n}$ 
 takes  the   factorized form \cite{Tseytlin:1984wj,Fradkin:1983zz,Deser:1983mm,Deser:1983tm}
\be
\td \OO_2 = \big(-D^{2}+\frac{1}{6}R\big) \big(-D^{2}+\frac{1}{3}R\big) \ ,  \la{35}
\ee
or $\td \OO_2 = (-D^{2}+2 ) (-D^{2}+ 4)$ for a unit-radius $S^4$  with $R=12$.  

To analyse the non-Einstein 
case of $S^{1}\times S^{3}$  background, let us    split  the   components of  $h_{\mu\nu}$ 
into $h_{ij}, \, h_{0i}, \, h_{00}$ 
and use that here $R_{00}=0$, $R_{ij} = \frac{1}{3} R\,g_{ij}, \ \ R=6$. 
Then,   $h_{ij}$
decouples  from $h_{0i}$ and $h_{00}$ in \rf{33}, with the 
transverse traceless  $h_{ij}$ dependent  part being 
\be
\label{TT}
\mathscr L^{(2)}_{S^1\times S^3} =\frac{1}{4}D^{2}h_{ij}\,D^{2}h^{ij}
-\frac{1}{3} R \,h_{ij}\,\partial_{0}^{2}\,h^{ij}-\frac{1}{6} R \,h_{ij}\,\mathbf{D}^{2}\,h^{ij}
+\frac{1}{36}\,R^{2}\,h_{ij}h^{ij}  \ ,  
\ee
where we used the notation in \rf{255}.
The corresponding 4-th order operator is thus 
\be 
\OO_2 =( \del_0^2 + \bD^2)^2 
-\frac{2}{3} R\,  (2\partial_{0}^{2} + \mathbf{D}^{2} )
+\frac{1}{9}R^{2}\ . \la{37}\ee
It is useful to rederive this expression   in the   2nd-derivative formulation 
of conformal higher spin theory involving auxiliary fields \ci{Metsaev:2007fq,Metsaev:2007rw}.
In the spin 2 case the  corresponding Lagrangian may be written as  \cite{Kaku:1977pa,Metsaev:2007fq}
\be
\mathscr L (g, f) = \sqrt g \big[ - f^{\mu\nu}\,G_{\mu\nu}-\frac{1}{4}\, f^{\mu\nu}\, f_{\mu\nu}+\frac{1}{4}\,
( g^{\m\n} f_{\m\n})^{2}\big] \ , \qquad\ \ \
G_{\mu\nu}\equiv  R_{\mu\nu}-\frac{1}{2}\,g_{\mu\nu}\,R\ . \la{38} 
\ee
Solving for the  auxiliary symmetric tensor   $f_{\mu\nu}$   we get back to the Weyl action \rf{31}. 
Expanding around a generic  curved background for $g_{\m\n}$   with 
$f_{\mu\nu}=-2 (R_{\mu\nu} - { 1 \ov 6} g_{\mu\nu} R)$  
it is straightforward  to find the  corresponding  quadratic fluctuation action  for 
$\delta f_{\m\n} = \p_{\m\n}$ and $ \delta g_{\m\n} = h_{\m\n}$.
Assuming  gauge conditions of  transversality and tracelessness of $h_{\m\n}$
and ignoring terms  involving Weyl tensor and  derivatives  of the curvature (as we are interested 
in a  conformally flat  constant-curvature background), we find\,\foot{The case of a 
  generic Einstein background  where the  fluctuation operator also factorizes  is discussed in  appendix \ref{A:Einstein}.}  
\be
\mathscr L^{(2)} = \mathscr L^{(2)}_{\phi\phi}+ \mathscr L^{(2)}_{\phi\,h}+
\mathscr L^{(2)}_{hh} \ , \la{310}
\ee
\be
\mathscr L^{(2)}_{\phi\phi} = - \frac{1}{4} \phi_{\alpha\beta} \phi^{\alpha\beta}
\ , \qquad 
\mathscr L^{(2)}_{\phi\,h} = \frac{1}{6} R\, h^{\alpha\beta} \phi_{\alpha\beta} -  \phi^{\alpha\beta} 
D_{\gamma}D_{\beta}h_{\alpha}{}^{\gamma}{}+ \frac{1}{2} \phi^{\alpha\beta} 
D^{2}h_{\alpha\beta} \ , 
\ee
\begin{align}
\mathscr L^{(2)}_{hh} =  & -\frac{1}{4}(R_{\gamma\delta}\,R^{\gamma\delta}-\frac{1}{3}\,R^{2}) \,h^{\alpha\beta}\,h_{\alpha\beta} 
+  2 R_{\alpha\gamma} R_{\beta\delta} h^{\alpha\beta} h^{\gamma\delta} 
+  R_{\alpha\beta} R_{\gamma\delta} h^{\alpha\beta} h^{\gamma\delta} \no  \\
& - 3 
 R_{\beta}{}^{\delta} R_{\gamma\delta}h_{\alpha}{}^{\gamma} h^{\alpha\beta}  +  R_{\gamma\delta} R^{\gamma\delta} h_{\alpha\beta} h^{\alpha\beta}
 + 3 R_{\beta\gamma} R\,  h_{\alpha}{}^{\gamma} h^{\alpha\beta}  -  \frac{19}{36}  R^2\,  h_{\alpha\beta} h^{\alpha\beta} \no  \\ 
& + \frac{1}{2} R^{\alpha\beta} D_{\alpha}\,h^{\gamma\delta} D_{\beta}\,h_{\gamma\delta} + R^{\gamma\delta} \, h^{\alpha\beta} 
D_{\beta}\,D_{\alpha}\,h_{\gamma\delta} -  \frac{5}{6} R \, h^{\alpha\beta} D^{2}\,h_{\alpha\beta} + 
\frac{1}{3} R\,  D_{\beta}\,h_{\alpha\gamma} D^{\gamma}\,h^{\alpha\beta} \no \\ 
& -  \frac{1}{2} R D_{\gamma}\,h_{\alpha\beta} D^{\gamma}h^{\alpha\beta} - 2 R^{\gamma\delta} \, h^{\alpha\beta} 
D_{\delta}\,D_{\beta}\,h_{\alpha\gamma} + R^{\gamma\delta}\, h^{\alpha\beta}  D_{\delta}\,D_{\gamma}\,h_{\alpha\beta} + 2R_{\alpha}{}^{\gamma}\,  h^{\alpha\beta} 
 D^{2}h_{\beta\gamma} \no  \\
& -  R^{\alpha\beta}\, D_{\gamma}h_{\beta\delta} \
D^{\delta}h_{\alpha}{}^{\gamma} 
+ R^{\alpha\beta}\,  D_{\delta}h_{\beta\gamma} D^{\delta} h_{\alpha}{}^{\gamma}
\  , \la{312}
\end{align}
where the first  term in \rf{312} comes from the expansion of the $\sqrt g$  factor  in \rf{38}.
Specializing to the case of the $S^{1}\times S^{3}$ background and concentrating on the 
part of the action that depends on the transverse traceless   spatial parts of the fluctuations $h_{ij},\p_{ij}$,  we get 
(after commuting derivatives, integrating by parts and using that $R_{00}=0, \ R_{ij}=\frac{1}{3} R\,g_{ij}$)
\iffa Here, the case of interest is $S^{1}\times S^{3}$ that we now analyse in details.
Again, we work in the temporal-like gauge 
\be
\label{gauge}
h_{00} = h_{0i} = 0.
\ee
We shall discuss the role of this gauge fixing when presenting the full partition function in 
section \ref{sec:spin2-Z}. 
We begin with $\mathscr L^{(2)}_{\phi\,h}$.
Commuting the covariant derivative in order to exploit transversality, we find
\be
\mathscr L^{(2)}_{\phi\,h} =\frac{1}{3} h^{\alpha\beta} R \phi_{\alpha\beta} - 2 h^{\alpha\beta} R_{\alpha}{}^{\gamma} \phi_{\beta\gamma} + \frac{1}{2} \phi^{\alpha\beta} D^{2}\,h_{\alpha\beta}.
\ee
We can restrict $\phi_{\alpha\beta}$ and $h_{\alpha\beta}$ to the {\bf purely spatial part} and set $R_{ij}=\frac{R}{3}\,g_{ij}$. This gives,
with $R=6$,
\fi
\be
\mathscr L^{(2)}_{\phi\,h} 
= \frac{1}{2}\,\phi^{ij}\,(D^{2}- { 2 \ov 3 } R )\,h_{ij} \ ,  
\qquad \qquad
 \mathscr L^{(2)}_{hh} = { 1 \ov 6} R \,  h^{ij}\,(\mathbf{D}^{2}-{ 1 \ov 2}  R )\,h_{ij} \ , \la{3144} 
\ee
so that \rf{310} becomes 
\be\la{314}
\mathscr L^{(2)} =
\frac{1}{2}\,\phi^{ij}\,(D^{2}-{ 2 \ov 3} R )\,h_{ij}- \frac{1}{4} \phi_{ij} \phi^{ij} +  {1 \ov 6} R\,  h^{ij}\,(\mathbf{D}^{2}-\ha R )\,h_{ij} \ .
\ee
Note that the {\em kinetic} $h h$ term  is  absent  in the Lagrangian \rf{38} expanded   near flat  space  but it 
appears on a curved background.  Using that $R=6$  (for a unit-radius $S^3$),   Eq.~\rf{314}  can be written also as 
\be\la{3114}
\mathscr L^{(2)} =
\frac{1}{2}\,\phi^{ij}\,(D^{2}-{ 4}  )\,h_{ij}- \frac{1}{4} \phi_{ij} \phi^{ij} + \,  h^{ij}\,(\mathbf{D}^{2}-  3  )\,h_{ij} \ .
\ee
Solving for $\phi_{ij}$,   we then find that $\mathscr L^{(2)} = {1 \ov 4} h^{ij}\, \OO_2\,h_{ij} $  where 
\be 
\OO_2 = (D^{2}-4)^{2}+4\,(\mathbf{D}^{2}-3)= \partial_{0}^{4}+ 2\,\partial_{0}^{2}\,( \mathbf{D}^{2} -4)  + ( \bD^2 -2)^2 \ , 
\la{316}\ee
 is indeed equivalent to \rf{37}. 
Note that $\OO_2$  can be written in the following   factorized form 
\be \la{3166}
\OO_2= \big[ ( \del_0 - 1)^2  + \bD^2 - 3 \big]\, \big[  ( \del_0 + 1)^2  + \bD^2 - 3  \big] \ , 
\ee
which is the $S^1 \times S^3$ counterpart of \rf{35}  found  in $S^4$ or AdS$_4$  case (where $R=\pm 12$).

\iffa 
The equations of motion of $\phi_{ij}$ are
\be
\phi_{ij} = (D^{2}-4)\,h_{ij},
\ee
leading to 
\be
\begin{split}
\mathscr L^{(2)} =& \frac{1}{4}\,h_{ij}\,\left[(D^{2}-4)^{2}+4\,(\mathbf{D}^{2}-3)\right]\,h^{ij}  \\
=&  \frac{1}{4}\,h_{ij}\,\left[
\partial_{0}^{4}- 2\,\partial_{0}^{2}\,(- \mathbf{D}^{2} +4)\mathbf{D}^{4}  + (- \bD^2 +2)^2 
\right]\,h^{ij},
\end{split}
\ee
in agreement with (\ref{S1S3-fact}). 
\fi

\subsection{Partition function on $S^1 \times S^3$}
\label{sec:spin2-Z}
Like   in the vector field case in section (\ref{sec:spin0}),   the derivation of the one-loop partition function 
can be presented either in 4-d   covariant gauge  or in the time-like reparametrization  gauge $h_{00}=0, \ h_{0i}=0$
with  the Weyl gauge $h^\m_\m=0$ being then equivalent  to the  tracelessness  of $h_{ij}$. 
Splitting   $h_{ij}$ into the transverse   and  longitudinal   parts  (and taking into account  various  ghost and  Jacobian factors
 as in the vector field case  discussed in section 2.2),
we end up with the following simple expression for the spin 2 analog of \rf{213}:
\be
\label{3116}
Z_2 = \frac{1}{\big[ \det \OO_{2\, \perp}   \,   \det' \OO_{1\, \perp}   \big]^{1/2}} 
\ . 
\ee
Here   the spin 2  operator $ \OO_{2\,\perp}$   given in \rf{37},\rf{316}
  is defined on transverse  traceless  tensors $h_{ij}$    while the spin 1  one is the same as  in \rf{213}.\foot{Note that this combination of determinants  describes the   right  number of degrees of freedom a conformal graviton: 
  $(6-1-1) +2=6$.}
The vector determinant  defined on transverse vector $V_i$ 
originates   from the decomposition $h_{ij} \to h_{ij}^\perp  + D_i V_j + D_j V_i$, $D^i V_i=0$. 
The prime  indicates that the  lowest $n=0$ mode 
of the vector Laplacian on $S^3$   is to be dropped  since this mode satisfies $D_i V_j + D_j V_i=0$   and thus cannot appear from $h_{ij}$  (see  appendix \ref{A:zero2}). 

\iffa
Before presenting the full partition function for conformal spin 2 fields, let us summarise the logic
behind our derivation. The gauge condition (\ref{gauge}) gives a ghost determinant of the form 
$\left[(\det \partial_{0}^{2})\det D^{2}\right]^{1/2}$. Splitting $h_{ij}$ into traceless transverse part plus
longitudinal vector $V_{i}$ gives another factor $\det(D^{2})^{1/2}$. The kinetic term $h_{ij}D^{4}h^{ij}$
splits into the TT part plus a vector term of the form $D_{i}V_{j}D^{4}D^{i}V^{j}$. The extra $D^{4}$
determinant cancels against ghost and measure analogous factors. Thus we end with a vector Laplacian contribution
from $V_{i}$. All in all, this means that we have to add to the TT spin 2 contributions the partition function 
of a lower spin 1 component.
\fi
 
 The spectrum of $\OO_{1\,\perp}$ was already given in \rf{216}.  Using that 
 the spectrum of the   transverse traceless   rank 2 tensor Laplacian on $S^3$    can be read off 
 from \rf{a2}--\rf{a3}, {\em i.e.} 
 \be 
( -\bD^2)_{2\,\perp}\ : \qquad   \l_n= (n+2) (n+4)  -2  \ , \ \ \ \ \ \ \   \dd_n= 2 (n+1) (n+5) \ , \ \ \ \ \ \ \ 
n=0,1,2,... \ , \la{317}
 \ee
 we  conclude  that the eigenvalues of  $ \OO_{2\, \perp} $  in \rf{316}    can be written as 
  ($\del_0^2 \to - w^2$, cf. \rf{29},\rf{216}) 
\ba  \la{318}
 \OO_{2\,\perp}\ : \ \qquad   \l_{k,n}= & \ w^4 + 2 w^2 \big[ (n+2) (n+4)  +2\big]   + \big[(n+2) (n+4)\big]^2\no  \\
 = &\  \big[ w^2  + (n+2)^2\big]\,  \big[ w^2   + (n+4)^2 \big]   \  . 
 \end{align}
 This  simple result  is  related to the special  structure of $\OO_2$ in \rf{316},\rf{3166}:  
   its eigenvalues  factorize with the  spatial parts being 
 squares of the effective energies $\omega_n$ which are  linear in $n$  as in the conformal scalar \rf{29} and vector \rf{216} cases. 
 This is also a  consequence of the underlying   conformal  invariance  of the spin 2 theory. 
 
 \iffa 
The TT part can be found by taking (\ref{TT}) and, in $S^{3}$, using the results of appendix \ref{A:Laplace}).
This leads to the  {\bf factorized} operator
\be
\begin{split}
\label{S1S3-fact}
& (\partial_{0}^{2}+\mathbf{D}^{2})^{2}-8\,\partial_{0}^{2}-4\,\mathbf{D}^{2}+4 =  \\
& \partial_{0}^{4}+\mathbf{D}^{4}+2\,\partial_{0}^{2}\,\mathbf{D}^{2}-8\,\partial_{0}^{2}-4\,\mathbf{D}^{2}+4 =  \\
& (\partial_{0}^{2}-(n+2)(n+4)+2)^{2}-8\,\partial_{0}^{2}+4\,((n+2)(n+4)-2)+4 =  \\
&\qquad \left[\partial_{0}^{2}-(n+2)^{2}\right]\,\left[\partial_{0}^{2}-(n+4)^{2}\right].
\end{split}
\ee
\fi

As a result, the conformal spin 2  partition function takes the standard  form  \rf{22} 
where the canonical  partition function  is a combination of the  two terms  corresponding to the two factors in \rf{3116}
\be
\Z_{2} = \Z_{2,0}(q)+\Z_{1,1}(q) \ . \la{319}
\ee
The notation  $\Z_{s,r}$ means that the sum over $n$ starts  with $n=r$,  so that 
 the explict   expressions following from \rf{318} and  \rf{216},\rf{217}  are  
\ba
&\Z_{2,0}(q) = \sum_{n=0}^{\infty} 2\,(n+1)(n+5)\,(q^{n+2}+q^{n+4}), \ \la{320}\\
&\Z_{1,1}(q) = \sum_{n=1}^{\infty} 2\,(n+1)(n+3)\,q^{n+2} \ . \la{321}
\end{align}
Doing the sums gives finally 
\iffa  The explicit sums 
in the above contributions are 
\be
\Z_{2,0}(q) = \frac{2 q^2 (3 q-5) \left(q^2+1\right)}{(q-1)^3},  \qquad
 \\
\Z_{1,1}(q) = 
-\frac{2 q^3 \left(3 q^2-9 q+8\right)}{(q-1)^3}.
\ee \fi 
\be
\Z_{2} =   { 10 q^2  - 18 q^4  + 8 q^5 \ov (1-q)^4} =  \frac{2 q^2 \left(5+ 5 q - 4 q^2\right)}{(1-q)^3}\ . \la{322}
\ee

\subsection{Partition function from conformal   operator counting in $\mR^4$}
\label{sec:spin2CFT}

Let us now show that   the same expression \rf{322}   for the conformal spin 2 partition function can be found in the operator 
counting method, {\em i.e.}  by treating the linearized   Weyl gravity \rf{31}   as a CFT  in $\mR^4$ 
and counting  gauge-invariant conformal operators  built out of the  linearized Weyl tensor $C\sim \del \del h$
weighted with their conformal dimension  and  subtracting the contributions   of gauge identities and equations of motion. 

As in the  conformal scalar and $d=4$ vector examples   discussed in section 2.1, adding all possible derivatives 
introduces the  universal overall denominator factor $(1-q)^{4}$, so the main problem  is to determine the numerator. 
First, let us  count the  off-shell components of  $C_{\m_1\n_1\m_2\n_2}$  modulo gauge identities. 
$C$   has dimension 2 ($h$ has  dimension 0)   and 10  independent components
(transforming as the (2,2) representation  of $SO(4)$);  this gives   $10 q^2$  contribution. 
Adding  all possible derivatives to $C$   produces  overcounting since 
there are non-trivial  gauge identities that $C\sim \del \del h$   satisfies,  {\em i.e.} 
\be
\cB^{\m_{1} \m_{2}}\equiv 
\varepsilon^{\mu_{1}\nu_{1}\gamma_{1}\delta_{1}}\,
\varepsilon^{\mu_{2}\nu_{2}\gamma_{2}\delta_{2}}\,
\partial_{\n_{1}}\partial_{\n_{2}}\,
C_{\gamma_{1}\delta_{1} \gamma_{2}\delta_{2}}=0 \ , \la{323}
\ee
and their derivatives. $\cB^{\m\n}$   has dimension 4   and is  symmetric and traceless  with  9 components, {\em i.e.}    this 
requires subtracting the $9 q^4$ term. However, subtracting all the derivatives  of $\cB^{\m\n}$    would also overcount as 
$\cB^{\m\n}$   itself satisfies   the  identity  $\del_\m \cB^{\m\l}=0$  which has dimension 5   and 4 components;
this requires adding back $4 q^5$.  Thus, the {\em off-shell}  count of the  components of the Weyl tensor and its derivatives 
leads to the following contribution to the partition function \rf{21} 
\be\la{324}
\Z^{\rm off-shell}_{2} = 
\frac{10\,q^{2}-9\,q^{4}+4\,q^{5}}{(1-q)^{4}}.
\ee
It remains to subtract also  some of the descendant operators $\del ...\del  C$     that  vanish due to the equations  of motion for 
the dynamical field $h_{\m\n}$, {\em i.e.} 
\be 
B_{\m_1\m_2}\equiv \partial^{\nu_1}\partial^{\nu_2}C_{\m_1 \n_1 \m_2 \n_2 }=0 \  . \la{325}
\ee
The count of  the symmetric traceless   $B_{\m_1\m_2}$  is the same as for the  $\cB^{\m_{1} \m_{2}}$  above.
We need  to  subtract $9 q^4$ but also to add back $ 4 q^5$ to account for the identity $\del^{\m} B_{\m\l}=0$. 
Thus the contribution of the  equations of motion that should be  subtracted  from \rf{324}  is 
\be\la{326}
\Z^{\rm e.o.m.}_{2} =  \frac{9\,q^{4}- 4\,q^{5}}{(1-q)^{4}}\ . \ee
As a result, 
\be 
\Z_2 = \Z^{\rm  off-shell 
}_{2} - \Z^{\rm e.o.m.}_{2} =  { 10 q^2  -  2( 9 q^4  - 4 q^5 )\ov (1-q)^4}, \la{327} \ee
is indeed the same    as \rf{322}. 

Let   us note that as in the $s=1$ case in section 2.1 
 here  (i) the contributions of the equations of motion and  of the gauge identities 
are    the same, {\em i.e.} they just double, and (ii)  the count of the equations of motion contribution is  the same 
as the count of the conserved   traceless   rank $s$   ``current"  operator  of dimension $2+s$ (conformal stress tensor for $s=2$). 
As  we shall discuss  in the next  section,  these   features   generalize to any spin $s$  conformal field  case   in $d=4$.

\iffa 
In order to match $\Z_{2}$ by the CFT operator counting method,  we consider conformal gravity 
expanded around flat  space and count multiplicities of components of 
\be
C_{\alpha\beta\gamma\delta}, \quad \partial_{\rho}C_{\alpha\beta\gamma\delta}, \quad \partial_{\rho}\partial_{\lambda}C_{\alpha\beta\gamma\delta}, \dots,
\ee
taking into account the Bach equations of motions and their derivatives
\be
B_{\gamma\delta}=\partial^{\alpha}\partial^{\beta}C_{\alpha\gamma\beta\delta}=0, \quad\partial_{\rho}B_{\gamma\delta}=0, \quad
\partial_{\rho}\partial_{\lambda}B_{\gamma\delta}=0, \dots.
\ee
Here, $C_{\alpha\beta\gamma\delta}$ is the linearised Weyl tensor around Euclidean flat  space $W(\delta+h) = C(h)+\mc O(h^{2})$. 
Notice that 
\be
[h] = 0, \quad [C] = [\partial^{2}h]=2.
\ee
The explicit counting goes as follows:
\begin{enumerate}
\item[a)] Components of $\partial^{n}\,C$: the associated generating function is 
\be
\label{Z2C}
\Z_{2, C}(q) = 
\frac{10\,q^{2}-9\,q^{4}+4\,q^{5}}{(1-q)^{4}}.
\ee
Here, the factor $1/(1-q)^{4}$ adds 4 dimensional derivatives, as before.
The term $10\,q^{2}$ is the number of components of $C$, transforming as the $2\times 2$ Young tableau 
representation  of $SO(4)$. The subtraction $-9\,q^{4}$, removes removes Bianchi identities~\footnote{
\red{we don't have a proof or a reference that these are the correct generalized Bianchi identities !}
}
\be
\mathscr B^{\delta_{1} \delta_{2}}=
\varepsilon^{\mu_{1}\nu_{1}\gamma_{1}\delta_{1}}\,
\varepsilon^{\mu_{2}\nu_{2}\gamma_{2}\delta_{2}}\,
\partial_{\gamma_{1}}\partial_{\gamma_{2}}\,
C_{\mu_{1}\nu_{1} \mu_{2}\nu_{2}}=0.
\ee
Finally, the term $9\,q^{5}$, takes into account the subtraction of the quantities
$\partial_{\delta_{1}}\mathscr B^{\delta_{1}\delta_{2}}$ that are identically zero.

\item[b)] Equations of motion $\partial^{n}\,(\partial^{2}\cdot C)$:
We first notice that the counting is equivalent to that of 
the components of a spin $2$ current $J$ which has dimension 4 and obeys a conservation law $\partial\cdot J=0$. Then, the associated generating function is 
\be
\label{Z2EOM}
\Z_{2, \rm EOM}(q)= \frac{9\,q^4-4\, q^{5}}{(1-q)^{4}},
\ee
where the term $9\,q^{4}$ is the number of symmetric 
traceless tensors with two indices in $d=4$ and $-4\,q^{5}$ removes the vector $\partial\cdot J$.
\end{enumerate}
Subtracting (\ref{Z2EOM}) from (\ref{Z2C}) we obtain indeed (\ref{Z2}).
\fi 
%
%
%

\section{Partition function  of general    conformal higher spin   field    in $d=4$}

Let us now  use the explicit $s=1,2$  results  found above as a motivation 
for the form  of the conformal higher spin 
operator and the structure of the associated partition function on the  $S^1 \times S^3$    background.
We  will  then  demonstrate    the agreement of the  resulting canonical  partition function $\Z_s$   with 
the one found directly  by   counting  gauge-invariant conformal operators   corresponding to 
 the  linearized conformal  higher spin action \rf{1} in $\mR^4$.

Namely, we shall assume that the $2s$-derivative conformal spin $s$ operator, evaluated on the   $S^1 \times S^3$ 
background  and restricted   to  transverse traceless   spin  $s>0$   tensors with 3-dimensional indices  $\phi_{i_1...i_s}$,
takes the following form:
\begin{align}
\mbox{$s$=even} :& \qquad 
  \    \OO_{s} = \prod_{p=1}^{s}\big[(\partial_{0}+2p-s-1)^{2}+\mathbf{D}^{2}-s-1\big], \la{41}\\
\mbox{$s$=odd} :&\qquad   \ \OO_{s} = -  
  \prod_{p=-\frac{s-1}{2} }^{\frac{s-1}{2}}
\big[(\partial_{0}+2p)^{2}+\mathbf{D}^{2}-s-1\big]
\ .\la{42} 
\end{align}
For example,   for $s=1$  we get  $\OO_1$ in \rf{213}, for $s=2$  we find  $\OO_2$ in \rf{3166},  while  for $s=3$ 
eq.\rf{42} gives $\OO_3= \big( \partial_{0}^{2}+\mathbf{D}^{2}-4\big)
\big[(\partial_{0}+2)^{2}+\mathbf{D}^{2}-4\big]\big[(\partial_{0}-2)^{2}+\mathbf{D}^{2}-4\big]    $, etc. 
Eqs.~\rf{41} and \rf{42}  may be   viewed    is the  $S^1 \times S^3$   counterpart
  of the  factorized form  \cite{Tseytlin:2013jya,Metsaev:2014iwa,Nutma:2014pua} 
of the $\OO_s$ operator on $S^4$ or AdS$_{4}$ background.\footnote{
For spin $s\ge 2$, a   derivation of the factorization (\ref{41}),(\ref{42}) from the quadratic 
expansion of a curved space action, as in section \ref{sec:spin2} for $s=2$, 
is not possible    because a general form  of a CHS action in curved space 
is  not yet available   (though one should be able  to derive  it in a general  conformally-flat background 
as in the AdS  case in \ci{Metsaev:2014iwa,Nutma:2014pua}). 
Nevertheless, one may use a  heuristic approach   by trying to generalize   the 2nd-derivative formulation 
of conformal higher spin theory  in flat   background   developed in \ci{Metsaev:2007fq,Metsaev:2007rw} to $S^1 \times S^{3}$  background. For example, 
for spin $3$, one can 
assume that,   like in the $s=2$ case  \rf{3114}, 
  the (gauge-fixed)  Lagrangian  should be  a quadratic form in three auxiliary symmetric traceless fields $\phi^{(1,2,3)}_{ijk}$ whose 
coefficients are linear combinations of  covariant derivatives $\partial_{0}$ and $\mathbf{D}^{2}$ 
allowed by dimension counting. Requiring that such a  Lagrangian $\mathscr L^{(2)}_3$ 
leads to the factorized operator $\mc O_{s}$ does not fix it completely,  but   imposes many constraints.
One  natural  solution  is 
$\mathscr L^{(2)}_3  =   \phi^{(1)}\, (\partial_{0}^{2}+\mathbf{D}^2 - 4)\, \phi^{(3)} +  \phi^{(2)}\, (\partial_{0}^{2}+\mathbf{D}^2 +4 ) \,\phi^{(2)}    - \phi^{(2)}\,\phi^{(3)} -  16\,\phi^{(2)}\, (\partial_{0}^2 -1)\,\phi^{(1)}$.} 

Using that   the spectrum of  Laplacian $-\bD^2$  on $S^3$ is given by (see  \rf{a1}--\rf{a2})
\be 
(-\bD^2)_{s\,\perp}:\ \ \ \ \ \ \   \l_n= (n+s) (n+s+2) -s \ , \ \ \ \ \ \   \dd_n= 2 (n+1) (n + 2s +1) \ , \la{43} \ee 
we then find that   the eigenvalues of $\OO_{s\,\perp}$  may be written in a simple form   generalizing 
\rf{216} and \rf{318}  ($w= 2 \pi k \b^{-1}$)
\be 
\l_{k,n}= \prod_{p=1}^s  ( w^2 + \omega_{n,p}^2 )   \ , \ \ \ \ \ \ \ \ \ \ \     \omega_{n,p} = n + 2p  \ . \la{44}\ee
Our proposal   for the   corresponding partition function  generalizing    \rf{213},\rf{3116} is then 
\be 
Z_s = { 1 \ov  \big[ \prod^s_{p=1} \det'  \OO_{p\, \perp}\big]^{1/2}  } \ , \la{45} \ee 
where prime  in $ \det'  \OO_{p\, \perp} $ means that  the product  of the eigenvalues  should start with  $n= s-p$, 
{\em i.e.} first  $s-p$ modes are to be omitted.  As in the spin 2 case,  this is related  to the fact  that these modes do not appear in the transverse traceless  decomposition of  the symmetric traceless  tensor $\p_{i_1...i_s}$
(see appendix \ref{A:zero2}).

The   canonical partition function corresponding to \rf{45}   can then be written as 
\be
\Z_{s}  = \sum_{p=1}^{s} \Z_{p,s-p}  = \Z_{s,0} +\Z_{s-1,1}  +\cdots+\Z_{1,s-1}  \ , \la{46}
\ee 
where  $\Z_{p,s-p}$   is the contribution of spin $p$ operator in \rf{45} following from \rf{44} 
\ba 
\Z_{p,s-p} =& \sum_{n=s-p}^{\infty} 2\,(n+1)(n+2p+1)\, \sum_{r=1}^p q^{n+2r}\no \\  
=& { 2 q^{s + 2 - p } (1 - q^{2p})  [ 1  + q    + (s^2- p^2) (1 -  q)^2   +    2s (1 - q)  ] \over   (1 + q) (1 - q)^4  } \ . \la{47}
\end{align}
Performing the sum  in    \rf{46} we then finish with 
\begin{align}
\Z_{s}(q) &= \frac{2\, q^2}{(1-q)^{4}}\,\left[(s+1)^{2}\,(1-q^{s})-s^{2}\,(1-q^{s+1})\right]\no  \\
&= {
2 (2s +1) q^2   - 2 (s+1)^2 q^{s+2} + 2 s^2 q^{s+3}
\ov (1-q)^4 } \la{48}\ . 
\end{align}
This  generalizes the  above  $s=1$ \rf{24} and $s=2$ \rf{322}  expressions to  any $s>2$. 


\iffa 
The analysis in (\ref{sec:spin2CFT}) can be done for generic spin $s$. As a warm up, we can consider the 
spin 3 case for which (\ref{CHS-Z}) gives
\be
\label{Z3}
\Z_{3}(q) = \frac{2 q^2 \left(9 q^3-7 q^2-7 q-7\right)}{(q-1)^3}.
\ee
The equations of motion for the spin 3, dimension 2, Weyl tensor $C_{\mu_{1}\nu_{1}\mu_{2}\nu_{2}\mu_{3}\nu_{3}}$
read \cite{Marnelius:2008er}
\be
B_{\nu_{1}\nu_{2}\nu_{3}} = \partial^{\mu_{1}}\partial^{\mu_{2}}\partial^{\mu_{3}}\,C_{\mu_{1}\nu_{1}\mu_{2}\nu_{2}\mu_{3}\nu_{3}}=0.
\ee
We can now repeat the analysis of  section \ref{sec:spin2CFT}. Using the same notation, we have now
\be
\Z_{3, C}(q) = 
\frac{14\,q^{2}-16\,q^{5}+9\,q^{6}}{(1-q)^{4}},
\qquad 
Z_{3, \rm EOM}(q) = \frac{16\,q^5-9\, q^{6}}{(1-q)^{4}},
\ee
The difference $Z_{3, C}-Z_{3, \rm EOM}$ is indeed equal to  (\ref{Z3}).
A discussion of the zero modes for $s=3$, similar to $s=2$ can be found in App.~(\ref{A:zero3}).
\fi 

Let us now generalize  the discussion in section 3.3  to $s>2$   to 
demonstrate that the same  expression \rf{48}   follows   indeed  from the operator counting method.
The generalized  Weyl tensor $C_{\m_1\n_1....\m_s\n_s}$  in \rf{1},\rf{2} 
transforms in the $(s,s,0,\dots,0)$   representation of   $SO(d)$  corresponding to the rectangular 
 Young tableau   with  two rows and $s$ columns
\be\la{49}
\underbrace{\begin{picture}(100,30)
\put(0,5){\line(1,0){100}}
\put(0,15){\line(1,0){100}}
\put(0,25){\line(1,0){100}}
\put(0,5){\line(0,1){20}}
\put(10,5){\line(0,1){20}}
\put(20,5){\line(0,1){20}}
\put(30,5){\line(0,1){20}}
\put(40,5){\line(0,1){20}}
\put(70,5){\line(0,1){20}}
\put(80,5){\line(0,1){20}}
\put(90,5){\line(0,1){20}}
\put(100,5){\line(0,1){20}}
\put(55,10){\makebox(0,0){$\cdots$}}
\put(55,20){\makebox(0,0){$\cdots$}}
\put(105,7){\makebox(0,0)[l]{ }}
\end{picture}}_{\mbox{$s$}}
\ee
Its   dimension   
 is given by 
(see, e.g., \cite{King79})
\begin{align}
&\nn_{(s,s)} =  { (2 s  + d -4) (2 s + d -3) (2 s+d-2) (s + d - 5)! (s + d  - 4)! \over s!\,(s+1)!\,(d-2)! (d-4)! }, \la{410} \\
&\nn_{(s,s)}\big|_{d=4} = 2\,(2\,s+1) \ . \la{411}
\end{align}
Since $C_s$  has  dimension 2, its components     contribute $ 2(2\,s+1) q^2$ to the numerator  of $\Z_s$.  
The non-trivial gauge identities  that generalize \rf{323}  (and which are implicitly contained in the general discussion in 
  \ci{Vasiliev:2009ck})   are 
\be
 \cB^{\m_{1}\cdots \m_{s}}\equiv 
\varepsilon^{\mu_{1}\nu_{1}\gamma_{1}\delta_{1}}\,\cdots\,
\varepsilon^{\mu_{s}\nu_{s}\gamma_{s}\delta_{s}}\,
\partial_{\n_{1}}\cdots \partial_{ \n_{s}}\,
C_{\gamma_{1}\delta_{1}\cdots \gamma_{s}\delta_{s}}=0 \ , \ \ \ \ \ \ \ \ \ \      \del_{\m_1}  \cB^{\m_{1}\cdots \m_{s}} =0\ , 
 \la{412}
\ee
where $\cB^{\m_{1}\cdots \m_{s}}$ has dimension $ s+2 $ and 
 is the totally symmetric  traceless tensor in the $(s,0,...,0)$ representation of $SO(d)$
of dimension 
\be \la{413} 
\nn_s= ( 2 s + d-2)   { (s + d-3)! \ov (d-2)! \,s!} 
 \ , \ \ \ \ \ \ \ \ \ \
\nn_s\big|_{d=4}=(s+1)^{2} \ . \ee
 Thus, we need to subtract the term $ (s+1)^{2}  q^{s+2}$.  We should also  add  back 
$s^2 q^{s+3}$  to account for the conservation  identity satisfied by $\cB_s$, {\em i.e.}   to  
compensate for the fact that the descendants of $\cB_s$  containing dimension $s+3$  spin $s-1$   operator 
$\del^{\m_1}  \cB_{\m_1...\m_s}$
are identically zero.  
In total,  the generalization of the off-shell  gauge-invariant operator count  in \rf{324} reads
\be\la{414}
\Z^{\rm off-shell}_s =  {2 (2s+1) q^2 - (s+1)^2\, q^{s+2} +s^2\, q^{s+3}\ov (1-q)^4  } .
\ee
The equations of motion for the  conformal spin  $s$ field following from \rf{1}  that generalize the linearized Bach  equations 
\rf{325} are 
\be 
B_{\m_1...\m_s} \equiv \del^{\n_1} ...\del^{\n_s}  C_{\m_1\n_1...\m_s\n_s}   =0 \ , \ \ \ \ \ \qquad 
\del^{\m_1} B_{\m_1...\m_s}  =0 \ .  \la{415}
\ee
  $B_{\m_1...\m_s}$  has exactly the same properties  as  $\cB^{\m_{1}\cdots \m_{s}}$
so  its counting  goes analogously, {\em i.e.} the generalization of \rf{326} is found to be 
\be\la{416}
\Z^{\rm e.o.m.}_s =  {(s+1)^2\, q^{s+2} - s^2\, q^{s+3}\ov (1-q)^4  } .
\ee
Subtracting \rf{416} from \rf{414} as in \rf{327},   we get exactly the partition function in \rf{48}. 

This  
provides strong support  to  
our conjectures  for the structure \rf{41},\rf{42} of the $s > 2$ conformal higher spin operator
 on $S^1 \times S^3$. 
 An alternative derivation of $\Z_s$   based on group-theoretic
argument   will be
presented in section \ref{sec:fromAdSCFT}.


\section{Conformal spin $s$   partition function from    CFT$_d$/AdS$_{d+1}$ perspective}
\label{sec:fromAdSCFT}

Let us  now return to the discussion in the  Introduction and 
elaborate on the relations \rf{7},\rf{14}  between 
  the  conformal  spin $s$ partition function  and the  partition functions 
associated with   spin $s$  conformal  operators  in  the singlet sector of free 
  CFT in $d=4$ and  thus  also with  the one-loop partition  functions  of massless    spin $s$ gauge field in AdS$_5$.
We will  suggest a method to compute the shadow  partition function $\Z_{-\, s}$ that will allow a generalization to all even dimensions $d$ 
and  will thus   lead to the expression for  the conformal higher spin partition function  $\Z_s$  generalizing \rf{48}
to  $d > 4$. 

\def \rx {{\rm x}}

\subsection{$d=4$ case}

The counting of the conformal higher spin  equation of motion operators $B_s$ \rf{415}   discussed  above  
 is literally  identical to the counting  of spin $s$  conserved current operators  $J_s$  in the singlet sector 
 of free  scalar CFT$_d$ 
 leading to the identification of 
 $ \Z^{\rm e.o.m.}_s$  in \rf{416}   with $ \Z_{+\, s}$ in \rf{8}.\foot{The full large $N$ singlet-sector partition function in $d \geq 4$ is 
 $\sum_{s=0}^\infty  \Z_{+\, s}(q) = [\Z_{\rm c.s.}(q)]^2$   which is the same as the one-loop partition function 
 in the full  higher spin theory in thermal AdS$_{d+1}$ (see \ci{sy,Giombi:2014yra}).}
  The latter  should in turn  be equal to the one-loop partition 
 function   $\Z_{s}^{(+)}$  of a massless spin $s$ field in thermal AdS$_{d+1}$  with $S^1 \times S^{d-1}$ boundary. 
 Indeed,  the expression \rf{88}  or \rf{416}  is the same as found in \ci{Gopakumar:2011qs,Gupta:2012he} 
 using  the thermal quotient  analog of massless spin $s$ field heat kernel in AdS$_5$. 
 
 The  direct  computation of  massless spin $s$ partition function with  the  alternative 
 boundary conditions in AdS$_{d+1}$  that   should   give 
 $\Z_{s}^{(-)}$  appears to be non-trivial.  The final   expression   should match    $ \Z_{-\, s}$, {\em i.e.} 
 the partition function that   counts shadow operators  of spin $s$  and  dimension $\D_- = 2-s$ 
 in   the scalar CFT$_d$. 
 
 Below   we shall first   determine the expression for $ \Z_{-\, s}$ by a somewhat   heuristic approach 
 and  in the  next section  derive  it by more rigorous group-theoretic method. 
   The final $d=4$ expression will be the same as in \rf{414} 
 thus confirming the    identification $ \Z_{-\,  s}  = \Z^{\rm off-shell}_s$   between  the shadow 
 partition function   and  the  {\em off-shell} part of the conformal  higher spin partition function 
  announced in  \rf{14}. 
 
 As   discussed  in the Introduction, a naive  guess \rf{10} 
 for $ \Z_{-\,  s} $ is  obtained by  replacing the dimensions in the expression \rf{8} for 
 $ \Z_{+\,  s} $ with their shadow  counterparts, {\em i.e.}  in $d=4$ 
 \be
\widetilde \Z_{-\, s} (q) = { (s+1)^2\, q^{2-s}-s^2\, q^{1-s}\ov (1-q)^4}  = \Z_{+\,s}(1/q)\ .\la{51}
\ee
 That  may  correspond to the  analytic continuation in  physical and ghost dimensions ($\Delta\to  d-\Delta$) in the  massless  higher spin 
 heat kernel expression  for $\Z_{s}^{(+)}$    on the  AdS$_5$ side. 
 However, $\widetilde \Z_{-\, s} (q) $   containing   poles in $q$  cannot be the  final   answer. 
 
 In general,  the shadow field with 
  dimension $\D_{-}=2-s$ corresponds to a  conformal group  $SO(4,2) $ 
 representation which   is below the unitarity bound  that   may thus contain singular states and their associated submodules.
 From the AdS$_{5}$ side, in the case of the alternative   boundary conditions 
 there are additional  gauge transformations  allowed by non-normalizability (see \cite{Giombi:2013yva} and refs. there).
These are in  one-to-one  correspondence with the 
conformal Killing tensors 
 that belong to the $SO(4,2)$ representation  $(s-1,s-1,0)$  labelled  by  the 
  Young tableau that has two rows with $s-1$ columns.\foot{Below we  sometimes use 
  also the simplified notation $(s-1,s-1)\equiv (s-1,s-1,0, ..., 0)$ for the 2-row reprepresentation of $SO(2r)$.}
 Notice that this representation is finite dimensional and  may be also described as a representation of the 
 maximal compact real form $SO(6)$, with the same Young tableau.  
  The dimension of $(s-1,s-1,0)$ is given in  \rf{410}   with 
$d=6$ and $s \to s-1$, {\em i.e.} 
\be
\nn_{(s-1,s-1,0)}\big|_{_{d=6}}  = \frac{1}{12} s^2 (s+1)^2 (2 s+1)\ .  \la{52}
\ee
This suggests  the  following 
form for the  full shadow partition function
(justified on a group-theoretic basis in appendix \ref{Verma})
\be
 \Z_{-\, s} (q) =\widetilde \Z_{-\, s} (q) +\sigma_{s}(q) \ , \la{53} 
\ee
where $\sigma_{s}(q)$ is the character   associated with the conformal algebra representation
 corresponding to the conformal Killing tensors.  It may be computed as 
a specialization of the $SO(6)$ character $\chi_{(s-1,s-1,0)}(\rx)$ (see appendix \ref{A:so-char}),
where we take a suitable limit of the parameters $\rx=(x_1,x_2,x_3)$  appropriate for 
counting the scaling  dimensions of the  states\,\foot{Here $\rx$ may be 
 associated with the   Cartan generators   of the compact subgroup $SO(2)\times SO(2)\times SO(2)\subset SO(4,2)$, {\em i.e.}, with 
  the scaling dimension  and two spins of $SO(4)$. 
}
namely,  $x_{1}=q$, and $x_{2},x_{3}\to 1$   (see  \rf{F22}).
 Explicitly,  we have\,\footnote{As recalled at the end of appendix \ref{A:so-char}, the character $\chi_{\ell}(\rx)$
 is a combination of determinants that turns out to be a polynomial in the components $x_{i}$ of $\rx$ and their inverses $x_{i}^{-1}$. So, the limit $x_{2},x_{3}\to 1$ is smooth. Alternatively, working with (\ref{E1}), one has to set 
   $x_{2}=x\to 1$ with $x_{3}=1$,
 as in (\ref{54}). 
 }
\be
 \sigma_{s}(q) = \lim_{x\to 1}\,   \chi_{(s-1,s-1,0)} (q, x, 1)
=\lim_{x\to 1}\,  { \det M (s-1; x,q) \ov \det M (0;x,q) }  \ , \la{54} \ee
 \be    M  (s-1; x, q)=
\left(
\begin{array}{ccc}
 2 & 2 & 1 \\
 x^{s+1}  + x^{-s-1} &x^s +  x^{-s} & 1 \\
 q^{s+1} + q^{-s-1}&q^s +  q^{-s} & 1 \\
\end{array}
\right)   \  , \qquad   M (0; x, q) = 
\left(
\begin{array}{ccc}
 2 & 2 & 1 \\
 x^2+ x^{-2} & x+  x^{-1} & 1 \\
 q^2+q^{-2} & q+  q^{-1} & 1 \\
\end{array}
\right)\ .\no \la{544}
\ee
The ratio of the  determinants in \rf{54}  can be  expressed as  
\be\la{56}
\begin{split}
\sigma_{s}(q) =& \ \frac{1}{6}\,s\,(s+1)\,(s^{2}+s+1) 
\\ & \  -\frac{1}{6}\,\sum_{p=1}^{s-1}p\,(p+1)\,\big[ (2s+1)p -3s^{2}-2s-1\big] \,
(q^{s-p}+q^{-s+ p})\ .
\end{split}
\ee
Doing the finite sum  and adding the result to \rf{51}, we  finally obtain for $\Z_{-\,s}$ in \rf{53} 
\be
\Z_{-\, s} = 
{ 2 (2s+1)\,q^{2}-(s+1)^{2}\,q^{s+2}+s^{2}\,q^{s+3}  \ov (1-q)^4}  \ . \la{57}
\ee
This is exactly the same as \rf{414}   and thus the relation $ \Z_{-\,  s}  = \Z^{\rm off-shell}_s$ 
is verified. 


\subsection{Generalization to $d> 4$}

The above   computation of $\Z_{-\, s}$   can be readily  generalized to  any even $d >4$. We 
start  with  the  known  expression for $\Z_{+\,s}=\Z^{(+)}_{s}$ in \rf{8}, {\em i.e.}
\be \la{888}
    \Z_{+\,  s}  =  { \nn_s\, q^{ s + d-2 }  - \nn_{s-1}\,   q^{s + d-1}   \ov (1-q)^d } \ , \ee
     that counts the   conserved spin $s$  dimension $\D_+ = s + d-2$ current operator  and its descendants.
The   {\em naive}  expression   for the shadow partition function  \rf{10}  is found by
 the $\D \to d- \D $ trick,  resulting in   $\wtd \Z_{-\, s}(q) =  \Z_{+\, s}(1/q)$.
The full  $\Z_{-\, s}$  
is obtained by correcting $\wtd \Z_{-\, s}$   by the character 
of the conformal group $SO(d,2)$ representation  associated with the conformal  Killing tensors, {\em i.e.} 
\be\la{58}
\Z
_{-\, s}(q) =  {{\nn}_{s}\,q^{2-s}-{\nn}_{s-1}\,q^{1-s} \ov (1-q)^{d}} +\sigma_{s}(q) \ , \ \ \ \ \ \ \ \ \ \ 
\sigma_{s}(q)= \chi_{(s-1,s-1,0,...,0)} (q, 1, ..., 1)  \ .
\ee
The   corresponding $\mso(d+2)$ character $\chi_{(s-1,s-1,0,...,0)} (q, 1, ..., 1) $   can be found  from (\ref{E1}),\rf{F22}, and the 
meaning of the choice of its arguments  is like in the previous $d=4$ case (see also the discussion at the end of 
appendix \ref{A:so-char}).
As a result,  we  find that $\Z
_{-\, s}(q) $ in \rf{58}   may be written in the following form  with only positive 
 powers of $q$ in the numerator 
\be\la{511} 
\Z_{-\, s}(q) = \frac{1}{(1-q)^{d}}\Big[\sum_{m=2}^{d-2} (-1)^{m}\,{\rm c}_{s, m}\,q^{m}-{\nn}_{s}\,q^{s+d-2}+{\nn}_{s-1}\,q^{s+d-1}\Big] \ . 
\ee
We may  thus  represent $\Z_{-\, s}$ as 
\be\la{5111} 
\Z_{-\, s}(q) = \hat \Z_s (q)  -  \Z_{+\, s}(q) \ , \ \ \ \ \ \ \ \ \ \ \ 
  \hat \Z_s \equiv   \frac{1}{(1-q)^{d}}\, \sum_{m=2}^{d-2} (-1)^{m}\,{\rm c}_{s, m}\,q^{m}  \ . 
\ee
Here the coefficients $\nn_s$ are the same as  in \rf{9},\rf{413}, while 
${\rm c}_{s,m}$  are some integers obeying ${\rm c}_{s,m}={\rm c}_{s,d-m}$  that are 
  readily found for  any  given value of $d$;   their 
 group theoretic interpretation as dimensions 
   of $(s,s,1,...,1,0,...,0)$  representations  of $SO(d)$   with $m-2$  rows of length 1
 will be discussed  in the next subsection.  
 For example, 
 in  $d=4$  we  have $\nn_s= (s+1)^2$     and (cf. \rf{57})
 \be   {\rm c}_{s,2}= 2 (2s+1)   \ . \la{515} \ee
 In   $d=6$,  where   ${\nn}_{s} =  \frac{1}{12} (s+1) (s+2)^2 (s+3)$ (see \rf{413}),   we find 
\be
{\rm c}_{s,2} = {\rm c}_{s,4} = \frac{1}{12} (s+1)^2 (s+2)^2 (2 s+3), \qquad
{\rm c}_{s,3} = \frac{1}{6} s (s+1) (s+2) (s+3) (2 s+3) \ . \la{5133}
\ee
As a result,   using  \rf{7},\rf{888},\rf{5111}, we
 arrive at the  following expression  for the  conformal higher spin partition function in any  even $d\geq 4$
\be
\begin{split}
\label{512}
\Z_{s}(q) &= \Z_{-\,s}(q)-\Z_{+\, s}(q) =    \hat \Z_s (q) - 2 \Z_{+\, s} (q) \\
&= \frac{1}{(1-q)^{d}}\bigg[\sum_{m=2}^{d-2} (-1)^{m}\,{\rm c}_{s, m}\,q^{m}
-2{\nn}_{s}\,q^{s +d-2}+2{\nn}_{s-1}\,q^{s + d-1}\bigg]\ . 
\end{split}
\ee
This  is a generalization of  the $d=4$ expression in \rf{11},\rf{48}. 
 
The explicit   form of the partition function \rf{512} in general $d$ is   readily  found 
 for few  low values of $s$. 
 The spin 0 case is special  in that there are no gauge redundancies. Only the first terms 
 in  $\Z_{-\,s}(q)$ \rf{511}  and $\Z_{+\, s}(q)$  \rf{8} are actually  present,  so the expression
 in \rf{512}   should be modified accordingly (the coefficient of the middle  term in the bracket
 is 1 not 2  and the last term is  absent). This  gives 
 \be \Z_{0}(q)  = \frac{ q^2   -  q^{d-2}  }{(1-q)^{d}} \  . \la{513}\ee
For  $s=1$  we get the following generalization of \rf{24}
\ba  &\Z_{+\, 1} =   \frac{1}{(1-q)^{d}}  ( d\, q^{d-1} - q^d  ) \ , \ \ \ \ \ \ \ \ \ \ \ \ 
\Z_{-\,1} = 
 \frac{1}{(1-q)^{d}}  \sum_{m=2}^{d}(-1)^{m}\binom{d}{m}q^{m} \ ,  \la{514}\\
&  \Z_{1}  = \Z_{-\,1}- \Z_{+\,1} =
 \frac{1}{(1-q)^{d}}\,\Big[ \sum_{m=2}^{d-2}(-1)^{m}\binom{d}{m}q^{m}  
 - 2d\, q^{d-1} + 2\, q^d \Big] \no \\
 &\qquad\qquad\qquad\ \ \ \ =   1- \frac{1-d\,q-q^{d}+d\,q^{d-1}}{(1-q)^{d}}
\ . 
\end{align}
For $s=2$  we find 
\be
\Z_{2}(q) = \frac{1}{2}\,\Big[ d^{2}+d+2+{(d-2)(d+1)}\,\frac{1-q^{d}}{(1-q)^{d}}\Big]-d\, \Big[
\frac{q^{d}+(1-q)^{d}}{(1-q)^{d-1}}-\,\frac{1-(1-q)^{1-d}}{q} \Big].\la{5155}
\ee
\iffa \begin{align} \begin{split} \Z_{2}(q) &= \frac{1}{(1-q)^{d}}\, \bigg[  q^d \left(-\frac{d^2}{2}+d    q-\frac{d}{2}+1\right)\\
   &+\left(\frac{d^2}{2}+d    q+\frac{d}{q}-\frac{d}{2}+1\right)    (1-q)^d+\frac{d^2}{2}-\frac{d}{q}+\frac{d}{2}-1\bigg]. \la{516} 
\end{split}
\end{align}\fi
These expressions can be checked by  direct  operator counting  in the $d \geq 4$ conformal higher spin theory  \rf{2}. 
For $s=0$  we have  $C_0= \del^0 \p_0 = \p_0$  of dimension 2 (see \rf{3}) 
with the equations of motion being $\del^{d-4} \p_0=0$; this  explains  the two terms in the numerator of \rf{513}. 
$\Z_0$ in \rf{513}  vanishes in $d=4$  and is equal to the standard 2-derivative conformal scalar partition function \rf{23} 
in $d=6$.

For $s=1$  the first term in the sum  in 
\rf{514}  is $\binom{d}{2}q^{2} = \ha d (d-1) q^2$ 
which is the contribution of components  of the dimension 2  field strength $C_1 = (F_{\m\n})$
of the  conformal vector in $d$ dimensions  with Lagrangian  $F^{\m\n} \del^{d-4} F_{\m\n}$,  (see  \rf{2}). 
The second term  $-\binom{d}{3}q^3 $  subtracts   the contribution of the 
 identity
$\cB^{\m_1...\m_{d-3}}=\epsilon^{\m_{1}...\m_{d}}\,\partial_{\m_{d-2}}\,F_{\m_{d-1}\m_{d}}=0$
which has indeed $\binom{d}{d-3}= \binom{d}{3}$   components. 
This  overcounts as  derivative  of $\cB^{\m_1...\m_{d-3}}$ vanishes identically, 
so   we need an extra  terms to compensate for  this, etc., etc.  This explains the presence of all $d$ terms in the sum 
in $\Z_{-\,1}$ in \rf{514}. The first term in the numerator  in 
$\Z_{+\,1}$ is the  contribution of the equations of motion $B_\m = \del^\n \del^{d-4} F_{\m\n}=0$  where $B_\m$ has $d$ components and dimension $d-1$. The second term accounts for the identity $\del^\m B_\m=0$. 

Similar  counting   can be repeated  also   for $s=2$  but is more cumbersome.
These  counting  checks demonstrate consistency of  the general expressions \rf{58},\rf{511},\rf{512}. 

\iffa The $\Z_{+\,1}$ part
is given by (\ref{8}) and does not deserve comments. The expression (\ref{514}) is the same as 
\be
\la{516}
\Z_{-\,1}(q) = 1+\frac{d\,q-1}{(1-q)^{d}}.
\ee
The counting interpretation of this expression is readable after writing
\be
\la{517}
\begin{split}
\Z_{-\,1}(q) &= \frac{1}{(1-q)^{d}}\,\sum_{k=2}^{d}(-1)^{k}\binom{d}{k}q^{k} \\
& = 
 \frac{1}{(1-q)^{d}}\,\bigg[\binom{d}{2}\,q^{2}-\binom{d}{3}\,q^{3}+\binom{d}{4}\,q^{4}+\cdots-d\,q^{d-1}+q^{d}\bigg].
\end{split}
\ee
\fi

\iffa {\color{red}{
--------------------------------------------------------------------------------------------
\subsection{Double trace deformation ( to be fixed and moved...)}
Another route to obtain $\Z_{s}$ is based on the double-trace deformation 
arguments of \cite{Giombi:2013yva}. Let us review the main idea.
To begin, let us consider the coupling of the $D=4$
conformal  $\mathcal N=4$  SYM theory  to a  background
conformal supergravity multiplet. Integrating out the SYM fields
one finds an  induced  action  for the conformal supergravity fields  
\cite{Fradkin:1983tg,Liu:1998bu,Buchbinder:2012uh}:
$ S_{\rm eff} \sim   \int  C_{mnkl}   \ln (L^{-2} \nabla^2 )  C_{mnkl} + ...  \sim   \int  (  C^2_{mnkl} + ...)+ ${ non-local} terms.
This relation  can be generalized \cite{Tseytlin:2002gz}  by  starting   with the  free  $\mathcal N=4$ gauge  theory   and coupling it to a higher spin  generalization of the conformal supergravity multiplet. For instance, 
let us consider the  bosonic  conserved traceless bilinear currents 
  $J_{m_1...m_s}\sim  X_r \partial_{n_1} ... \partial_{n_s} P^{n_1...n_s}_{m_1 ...m_s}   X_r $  ( $X_r$ stand for the CFT fields)
  of    dimension $\Delta= 2 + s$.
 Coupling them  to  a    background higher spin conformal  field $\phi_s$, 
integrating out the free SYM fields ({\em i.e.} computing the logarithm  of the determinants in the background)
 and expanding the resulting induced effective action  for $\phi_s$
 to quadratic order  one  then gets  the logarithmically divergent term proportional to the
 CHS   Lagrangian, schematically  $\int d^4 x\ \phi_s P_s  \partial^{2s} \phi_s$. 
These remarks show that it is possible to write
\fi 


\subsection{Partition functions   from  characters  of  conformal  algebra representations}
\label{sec:fromBGG}

Our main  result (\ref{512}) for the conformal higher spin partition function
 can be understood at a deeper level in terms of  characters of
 Verma modules of the conformal algebra $\mso(d,2)$ (see appendix
\ref{A:BGG}  for details).
From a group-theoretic perspective, the  partition functions
$\Z_{+\,s}$ and $\Z_{-\,s}$   are
associated with the
conformal current and the shadow fields.
 The  important  point is that the conformal current and the shadow
field are not equivalent as $\mso(d, 2)$- modules.
  Instead, the conformal current generates a unitary irreducible module while the shadow field generates an indecomposable $\mso(d, 2)$-module which is reducible (the Weyl-tensor like  field strength built out of a shadow field is a conformal primary) 
  and non-unitarizable (its  dimension 
  $\Delta_-=2-s$ is below the unitarity bound) \cite{Bekaert:2012vt}. 
  The  analysis of the relevant    $\mso(d, 2)$-modules has been presented in full generality in 
\cite{Shaynkman:2004vu} where the corresponding  resolutions \`a la Bernstein-Gelfand-Gelfand (BGG) have been derived. 
We have specialised  it to our case in   appendix~ \ref{A:BGG}, explaining in some detail  various 
technical aspects. 

\def \rZ  {{\rm Z}}


The conformal  spin $s$ partition function $\Z_{s}(q)=\Z_{-\,s}(q)- \Z_{+\,s}(q)$   can be interpreted as {\em on-shell}   shadow 
field partition function, with  $\Z_{-\,s}$   being the {\em off-shell} one.  Eq. \rf{F19}, {\em i.e.} 
${\cal D}_{[2;(s,s)]}(q,\rx)={\cal S}_{[2-s;(s)]}(q,\rx)-{\cal D}_{[s+d-2;(s)]}(q,\rx)$, is  relating  the character corresponding to 
the dimension 2
conformal  field in  $(s,s)=(s,s,0,\dots, 0)$  representation of $\mso(d)$ to the  characters   of the 
shadow field of dimension 2  
and of the conserved current field  of dimension $s+d-2$  both  in $(s)=(s,0,\dots, 0)$  representations of $\mso(d)$.
The explicit expression for ${\cal D}_{[2;(s,s)]}(q,\rx)$ is given in eq.\rf{F32}. 

To specify \rf{F32} to the case of the partition function counting only scaling dimensions (with no  
{\em chemical potentials}  $x_i$ 
for  $\mso(d)$ spins)
we need to set  there  $\rx=(1, ...,1)$. 
Then,  according to \rf{f4},  $P(q,\rx) \to (1-q)^{-d}$. 
Also,   in this case  the $\mso(d)$ characters $\chi_\ell (\rx)$  (see \rf{E1})   that appear in  \rf{F32}  read 
$\chi_{\ell}(1,\ldots,1)$, {\em i.e.}  are given by   the dimension $ \mbox{dim}(\ell)$ of the 
corresponding finite-dimensional $\mso(d)$ representation     with  Young tableau $\ell=(\ell_1, ...,\ell_{d\ov 2})$.
The dimension of $(s)=(s,0,\dots,0)$  representation  of $\mso(d)$  is given by $\nn_s$ in \rf{9}, 
the dimension of $(s,s)=(s,s,0,\dots,0)$ is given in \rf{410},   while 
the  dimension of the representation $(s,s,1^{p})=(s,s,1,...,1,0, ...,0)$  associated with the  Young tableau
with two length $s$ rows and $p$ additional length 1 rows   can be found from the algorithm in 
\cite{King79}.

Then,  (\ref{F32})   takes   exactly the same form as   (\ref{511}),  with the coefficients ${\rm c}_{s,m}$ in \rf{511} 
now  having an explicit group-theoretic interpretation   as dimensions of  the corresponding  $\mso(d)$ representations
with two rows of length $s$ and $m-2$ rows of length 1:
\ba
{\rm c}_{s,m} =& \dim (s, s, 1^{m-2})\la{62}  \\
=& \ \frac{(2s+d-2)!(s+d-3)!(s+d-4)!(s+d-3-m)!(s+m-3)!}{(2s+d-5)!(s+m-1)!(s+d-1-m)!s!(s-1)! (d-2)!(d-2-m)!(m-2)!}
\ . \no
\end{align}
Let us note also that the partition function  corresponding   to the conformal Killing tensors,
given by \rf{F22} with $\rx=(1, ..., 1)$, 
is the same as $\sigma_s(q)$  expressed  in terms of the $\mso(d+2)$   character in \rf{58}. 

\def \del {\partial}

The above discussion applied to all  even dimensions $d\geq 4$. Let us now comment on the  special  case of $d=2$. 
In $d=2$   the classical   CHS action \rf{2} is trivial (the $s >1$ field strengths   vanish), so
this partition function comes only from gauge freedom,  {\em i.e.} from   gauge fixing terms  and 
ghosts in path integral approach (see also \ci{Tseytlin:2013fca}). 
Assuming $s >1$,   it  follows  from the general expression \rf{512}. For $d=2$  the first term in the bracket    is  absent and  
thus we get ($\nn_s=2$,  see \rf{413})  
\be
\Z_{s}(q) 
 = { - 4 q^s + 4 q^{s+1} \ov (1-q)^2 } =  - \frac{4\,q^s}{1-q} \ , \ \ \ \ \ \ \    s > 1   \ .  \la{516} \ee
 Equivalently, from \rf{888},\rf{58},\rf{5111},
 \be   
 \Z_s =\Z_{-\,s} -  \Z_{+\,s} =    - 2  \Z_{+\,s}  \  , \qquad \ \ \qquad   \hat \Z_s=\Z_{-\,s}   +  \Z_{+\,s} =0 \  , \ \ \ \ \     s > 1  \ .   \la{517} 
\ee
For $s=1$, {\em i.e.} the conformal   vector field  in $d=2$ with the non-local (Schwinger) action 
$\int d^2 x\,   F^{\m\n} \del^{-2} F_{\m\n}$,  the result is actually  half  of that  of \rf{516} with $s=1$.\foot{Here the counting of terms  in the numerator goes as 
follows: $F_{\m\n}$  of dimension 2  gives $q^2$, there are no gauge identities, while the equations of motion give  
$ \del^{-2} \del^\m F_{\m\n} =0, \ \del^{-2} \del^\m \del^\n F_{\m\n} =0$, {\em i.e.}   $2q -  q^2$.  In total, 
$\Z_1= {  q^2- 2q + q^2\ov (1-q)^2} = -{ 2q\ov 1-q}$.}
If we also   include dimension 2 scalar   with the $\int d^2 x\,    \phi\, \del^{-2} \phi$, then  a similar count gives  the numerator 
of $\Z_0$ as $q^2 - 1$, so that 
\be 
\Z_1 =  -{ 2q\ov 1-q}  \ , \ \ \ \ \ \ \ \ \  \quad   \Z_0  =  -{ 1 + q \ov 1-q} =  \Z_1 -1  \ .   \la{5117}\ee
Equivalently, specializing the discussion of appendix \ref{A:BGG} to the  $\mso(2,2)=\mso(2,1)\oplus \mso(2,1)$  case,one 
finds that   the modules ${\cal V}_{[2;(s,s)]}$ and thus ${\cal D}_{[2;(s,s)]}$ are 
absent from the BGG sequence. 
The partition functions (characters)  corresponding to the conserved current $[s;(s)]$, 
shadow $[2-s;(s)]$,  conformal Killing tensor $[1-s;(s-1)]$  and  the conformal  spin $s$ field   $[2;(s,s)]$
representations of $\mso(2,2)$  are then given, respectively,  
by
\be\la{65}
 \Z_{+\,s}= -2 C_{s-1}(q)\ ,\ \ \ \ \ \  \ \Z_{-\,s}=2\,C_{s-1}(q)\ ,\ \  \ \ \ \ \ \sigma_s= 2\chi_{s-1}(q) \ , \ \ \ \ \ \Z_s=   -4  C_{s-1}(q) \ , 
 \ee
where  $C_s(q)=\frac{q^s}{1-q^{-1}}$  is the character of the  $\mso(3)$  Verma module corresponding to a highest weight $s$  
representation.\foot{Here $s$ may  be any real number in general, see\cite{Dolan:2005wy}. 
$\chi_s (q)$ is the standard $\mso(3)$ character.  
Note also that 
$C_s(q^{-1})=-C_{-(s+1)}(q), \ \   \ C_{s-1}(q)+C_{s-1}(q^{-1})=\chi_{s-1}(q) $.}

\def \ep {\epsilon} 
\def \vep {\varepsilon}

\section{Summing over spins:  total CHS partion function and vanishing of Casimir energy}

Having found  the  conformal  higher spin partition function  for fixed $s$, {\em i.e.} \rf{48}  for  $d=4$  and \rf{512}  (with \rf{413},\rf{62})
for general $d$,  we are ready to  perform the sum over all spins $s=0,1,2,...,\infty$ 
to find to the  total CHS theory partition function. 

The sum of the $\Z_{+\,s}$ part \rf{8} of $ \Z_s =\Z_{-\,s} -  \Z_{+\,s}$ is finite (assuming $q < 1$) 
\be 
\Z_+ (q)  = \sum_{s=0}^\infty  \Z_{+\,s} =   { q^{d-2} (1-q^2)^2 \ov (1-q)^{2d}  }   \ . \la{71}
\ee
Since  $\Z_{+\,s} $  \rf{888} is the same as the massless  spin $s$ partition function in $AdS_{d+1}$,   Eq.\rf{71}  is 
equal to the partition   function of  totally symmetric  massless  higher spin Vasiliev   theory. It is   
 also the same as    leading order  term in the large $N$ limit of the 
canonical partition function of  the singlet sector of $U(N)$ scalar theory on $S^1 \times S^{d-1}$  \ci{Giombi:2014yra}
(equivalently, it counts all  spin $s$ conserved current operators  in free scalar CFT$_d$). 

The  sum of  $\hat \Z_s$  in \rf{5111} and thus of $\Z_{-\,s}$ in \rf{511}  is, 
however,   divergent:  the coefficients  ${\rm c}_{s,m}$  in $\hat \Z_s$ 
\rf{5111} are polynomials in $s$  (given  by \rf{62},\rf{71}), and  their  growth with $s$ is not suppressed 
 by $s$-independent powers of $q$.
 In the discussions of  the  conformal $a$-anomaly coefficient of CHS theory   in
  \ci{Giombi:2013yva,Tseytlin:2013jya,Tseytlin:2013fca,Giombi:2014iua} 
  it was noticed   that 
  there is a   natural   generalized $\zeta$-function or cutoff 
  regularization  \ci{Giombi:2014iua} that leads to the vanishing of the total  sum of the individual 
  anomaly coefficients over $s$:\foot{Here $\Big|_{\epsilon\to 0,\ \rm fin}$ means   keeping  only finite terms in the   sum in the limit $\ep\to 0$, 
  {\em i.e.}  dropping all  poles in $\ep$.}
  \ba
&\qquad \qquad \qquad  \qquad  \sum_{s=0}^{\infty} e^{- (s+\frac{d-3}{2})\, \ep}   \, {a}_{s} \Big|_{\epsilon\to 0,\ \rm fin} =0 \ ,  \  \ \ \ \ \ \la{72} \\
&   d=4: \ \ \ \ \  a_s= {1 \ov 180} \nu^2 ( 14 \nu + 3) \ ,  \ \ \ \ \ \qquad\qquad  \qquad \nu = s (s+1)   \  ; \la{73}\\
 &   d=6: \ \ \ \ \  a_s= {1 \ov 151200} \nu^2 (22 \nu^{3}  - {55} \nu^2  -2 \nu + 2) \ , \ \quad \nu = (s+1) (s+2)  \ ; \ \ \ \ \ \  ... \  \ . \la{74}
  \end{align}
  This regularization should be consistent with the underlying symmetries of the CHS 
  theory  and so it  is an obvious   prescription   to 
   to define the total partition function. 
  Then   $\Z  =  \sum_{s=0}^{\infty} e^{- (s + {d-3\ov 2})\, \ep} \, \Z_{s} \Big|_{\epsilon\to 0,\ \rm fin}$ is given by 
 \ba
&  \qquad \qquad  \qquad \qquad  \qquad \qquad \Z(q) = \hat \Z(q)   - 2 \Z_+ (q) \, 
 \ , \ \  \la{75}\\
& \hat \Z(q) =       \frac{1}{(1-q)^{d}}\, \sum_{m=2}^{d-2} (-1)^{m}\, \hat  {\rm c}_m \, q^m  \   ,  \ \ \ \ \ \   \  \qquad
\hat  {\rm c}_m \equiv  \sum_{s=0}^{\infty} e^{-(s+\frac{d-3}{2})\, \ep}   \, {\rm c}_{s,m}\, \Big|_{\epsilon\to 0,\ \rm fin}\ .\la{76} 
\end{align}
Using \rf{515},\rf{5133},  we thus find  for $\Z=\sum_s \Z_s$
\ba 
&  d=4: \ \ \ \ \  \Z= -\frac{q^2(11 +26 q+11q^2)}{6 (1-q)^6}  \  , \la{77}\\
 &   d=6: \ \ \  \Z=  \frac{q^2 (407 -5298 q-466311 q^2-992956 q^3-466311 q^4-5298
   q^5+407 q^6  )}{241920 (1-q)^{10}} . \no  \la{78}
  \end{align}
In general, the summed expression  has  the same symmetry  property  as $\Z_+$ in \rf{71}, {\em i.e.} 
\be
\Z(q) = \Z(1/q) \ . \la{78}
\ee
Given the canonical partition function $\Z(q)$ \rf{21}, 
the corresponding Casimir energy  on $S^{d-1}$  can be found  from the standard relations 
(see, e.g.,   \ci{gpp}, cf. \rf{21} ) 
\be 
E_c= \ha \sum_n  \dd_n\, {\om_n}  = {1\ov 2} \zeta_E (-1) \ , \ \ \ \ \ \ \ \ \ \ \ 
 \zeta_E (z) = {1\ov \Gamma(z) } \int^\infty _0 d \beta \, \beta^{z-1} \, \Z(e^{-\b}) \ .  \la{79}
\ee
Then, by the same argument as in the  case of the massless  higher spin partition function $\Z_+(q)$ 
in \ci{Giombi:2014yra}  it follows from \rf{78}   that the total summed over $s$ 
Casimir energy on $S^{d-1}$   in the conformal higher spin theory also vanishes. 

It is  of interest to derive this also directly from 
 the explicit  expressions for the  Casimir energies of individual  conformal spin fields. 
Let us  start with the  $d=4$ case. To compute the corresponding Casimir energy $E_{c, s}$ one may either use \rf{79} 
with $\Z_s$   given by \rf{48}, or  write $\Z_s$  as 
(noting  that  $(1-q)^{-4}= \sum_{n=0}^{\infty}\binom{n+3}{3}\,q^{n}$)
\be
\Z_{s} = \sum_{n=0}^{\infty}\frac{1}{6} (n+1) (n+2) (n+3) \Big[2 (2 s+1)
   q^{n+2}-2 (s+1)^2 q^{n+s+2} + 2 s^2 q^{n+s+3} \Big] \ , \la{710}
\ee
and extract the corresponding degeneracies  and  mode  energies  to   get $E_{c,s} =\sum_a  \sum_n \dd^{(a)}_n \omega^{(a)}_n $. 
Then,  using the    spectral  cutoff  or spectral $\zeta$-function  regularization as standard  for the Casimir energy 
 (see also \ci{Giombi:2014yra}),
we  find
\ba
& E_{c, s} = \sum_{n=0}^{\infty}\frac{1}{12} (n+1) (n+2) (n+3) \Big[  2  (2  s+1)(n+2)\, e^{-(n+2) \ep } \no  \\
   &\ \  \qquad -2 (s+1)^2 (n+s+2)\, e^{-(n+s+2)\ep } + 2 s^2 (n+s+3)\, e^{-(n+s+3)\ep }    \Big]_{\ep\to 0\, ,\  \rm fin} \ , \la{711}
\end{align}
so that 
\be 
 d=4: \ \ \ \quad  \qquad  E_{c, s}= {1 \ov 720} \nu\, ( 18 \nu^2 -14 \nu -11) \ ,  \ \ \ \ \   \qquad \nu = s (s+1)   \  . \la{733}\ee
Note that this  expression is similar but different from \rf{73}  (in general, in $d>2$ the Casimir energy  and  the $a$-anomaly coefficients need not match, cf. the discussion in \ci{Giombi:2014yra}).  Still, the sum  of $ E_{c, s}$  also  vanishes 
when  computed with the same regularization as in \rf{72}:
\be 
\sum_{s=0}^{\infty} e^{- \, (s+\frac{d-3}{2})\, \ep}   \, {E}_{c,s} \Big|_{\epsilon\to 0,\ \rm fin} =0   \ . \la{743} \ee
The same  is true for any $d$.\foot{It is possible to check that 
the regularization with $e^{- (s+a)\ep }  $  leads to vanishing  sum over spins  for any $d$ only if 
$a= \frac{d-3}{2}.$}
For example,   in $d=6$  we find  the following explicit expressions for $\Z_s$ in  \rf{512}  and the corresponding 
Casimir energy  (cf. \rf{74})
\ba
& \Z_{s} = \frac{(s+1)(s+2) }{12(1-q)^{6}}\,\Big[(s+1) (s+2) (2 s+3)\, q^2   -  2 s  
   (s+3) (2 s+3)\, q^3\,  \,  \no \\ 
& \qquad \ \ \ \  +  (s+1) (s+2) (2 s+3) \, q^4 -2  (s+2) (s+3) \, q^{s+4}+2 s (s+1) 
    \, q^{s+5} \Big]\ ,\la{7444} \\
& d=6: \ \ \ \    E_{c, s}= \frac{1}{241920}\,\nu^{2}\,(12 \nu ^3-58 \nu ^2-6 \nu
   +117)  , \ \ \ \ \ \nu= (s+1) (s+2) \ ,  \la{744}
\end{align}
 and one can check directly   that  \rf{743} is satisfied.

 
 We remark that a different regularization 
 of the sum over spins than used in \rf{72} 
 would have led to a partition function for which $\Z(q)\neq \Z(1/q)$ and  as a result 
 the  corresponding 
 vacuum energy would have been non-vanishing.\foot{The  regularization  \rf{72},\rf{743} 
 is indeed a   very natural one as one can   show that it is a direct analog of the spectral regularization  used in \ci{Giombi:2014iua}.}
 So, having total $E_{c}=0$ and having right regularization
 are correlated facts (see  a related  recent discussion in \cite{Basar:2014mha}).
 

Finally, let us consider  the special case of $d=2$ \rf{516},\rf{517}.  In this  case $\Z_s$ is $-2\Z_{+\,s}$, 
{\em i.e.} is  -2   of the massless   spin $s$ partition function in AdS$_3$, 
and we get from \rf{516},\rf{5117}  the following expression for the summed   partition function 
\be \la{dd2} 
d=2: \ \ \ \ \ \ \ \ \ \quad    \Z= \Z_0 + \Z_1 + \sum_{s=2}^\infty \Z_s = - { (1+ q)^2 \ov (1-q)^2} \ , \ee 
so that once again $
\Z(q) =\Z(1/q) $ and the summed  Casimir energy vanishes. 
This can be seen also  explicitly  using that $\Z_s=-2\Z_{+\,s}$
implies that 
the corresponding   Casimir energies   are also proportional in the same way, {\em i.e.}    (cf. 
 eq. (5.31) in  \cite{Giombi:2014yra})
\be\la{616}
d=2: \ \ \ \ \ \ \ \quad      E_{c,s} = \frac{1}{6}\,\big[1+6\,s\,(s-1)\big]\  ,\qquad\ \ \ \  s>1 \ . 
\ee 
 In the special   cases of   ``non-local''  $s=0,1$   fields 
 we get  from \rf{5117}  half the $s=0,1$ extrapolated values of \rf{616}, {\em i.e.} 
$E_{c,0} =E_{c,1}= { 1 \ov 12}$. 
Using  the universal 
 regularisation (\ref{743})  to define the sum over $s>1$, we conclude
  once  again  that the  Casimir  energy
 of the full  CHS   theory  vanishes,\footnote{
This also follows from the fact that the total partition function is the function 
$\Z= \Z_0 + \Z_1   + \sum_{s=2}^{\infty}  \Z_s  = - (1+q)^2/(1-q)^2$. Its symmetry under  $q\to 1/q$
explains the vanishing of the Casimir energy.
 }
 \be 
d=2: \ \ \ \ \ \  E_{c,0} + E_{c,1}   + \sum_{s=2}^\infty e^{- \, (s-\frac{1}{2})\, \ep}   \, {E}_{c,s} \Big|_{\epsilon\to 0,\ \rm fin} =0 \ . \la{6177}
\ee
In $d=2$ the Casimir energy on $S^1$  is indeed proportional to the corresponding  conformal anomaly 
coefficient (central charge)  which thus also cancels.\foot{The Seeley coefficient  for a conformal spin $s>1$ field is  
$B_2 = \frac{1}{3} c = -\frac{2}{3} \,\left[1 + 6 \,s\,(s-1)\right] = - 4\, E_{c,s}$, see  eq. (A.7)  \cite{Tseytlin:2013fca}. 
The same  values 
 follow  \ci{Giombi:2013yva} also from the AdS$_3$  perspective, {\em i.e.} \rf{6}.
 For $s=1$   one finds $B_2 = \frac{1}{3} c =  -\frac{1}{3}$ \cite{Tseytlin:2013fca}.
 For the non-local scalar $\int \phi \del^{-2} \phi$ the central   charge is   minus  the value ($c=1$)  of the  real scalar  one, {\em i.e.} 
 once again  $B_2 =  -\frac{1}{3}$.}

\section{Concluding remarks}

In this paper   we have   found   the partition function $\Z$  of  non-interacting    conformal  higher spin (CHS) theory  viewed as a 
 collection of    free   spin $s$  CFT's  in $\mR^d$.  The same   partition function can be  computed from 
 the CHS  theory defined on a curved conformally flat  background $S^1 \times S^{d-1}$. We discussed  relations  to partition functions appearing in the context of AdS/CFT 
 and gave a   representation-theoretic interpretation of  $\Z$  in terms of  characters of conformal algebra $\mso(d,2)$.

While we  focused  on the   free CHS theory, we  should stress that  there  exists 
a full nonlinear  generalization. 
As is well known, 
introducing consistent interactions involving massless higher-spin fields is notoriously difficult and various no-go theorems
 express the incompatibility between higher-spin gauge symmetries
and minimal coupling with gravity around a flat background (see,  e.g.,  \cite{Bekaert:2010hw} for a review). 
Fradkin and Vasiliev discovered \cite{Fradkin:1987ks} that this incompatibility can be resolved on 
a constant curvature  (A)dS background,
and that  eventually lead Vasiliev to  his  unfolded equations describing an 
interacting tower of massless higher-spin fields \cite{Vasiliev:1990en,Vasiliev:2003ev}.
Only two other explicit examples of 
 interacting higher-spin theories  (in $d\geqslant 3$) are known at present:
Chern-Simons-like  theory  in $d=3$\,\footnote{The literature on this subject is considerable so we simply refer to 
\cite{Afshar:2013vka,Gonzalez:2013oaa}
for an example of CS theory in $d=3$ flat spacetime.}
 and conformal higher spin theory in even $d$ dimensions. 
In contradistinction with massless  higher spin  theory, these two  share the following attractive features:
their interactions are consistent with  
coupling to gravity even around a flat background and they admit a conventional action principle.

Indeed,  the non-linear CHS   theory can be   defined as  an  induced theory \ci{Tseytlin:2002gz,Segal:2002gd,Bekaert:2010ky,Giombi:2013yva}.
 As mentioned in the Introduction,  the logarithmically divergent piece of the one-loop  effective action 
of a   scalar CFT   coupled to   {\em source}  fields  $\p_s$    for all conserved   symmetric  spin $s$   currents
is   a  local functional   of $\p_s$  starting with the  CHS   kinetic term in \rf{2}\,\footnote{More precisely, the vertex 
involving a product of $m$ fields with spins $s_i$ ($i=1,\ldots,m$)
contains a total number $
p = d +\sum_{i=1}^m\,(s_i-2)
 $ of partial derivatives
  \cite{Bekaert:2010ky}. 
As one can see, the spin-two fields do not contribute in the sum, consistently with minimal coupling to gravity ({\em i.e.} spin 2 fields essentially enter through
general covariantization and do not modify the total number of derivatives).} 
and  can  thus be interpreted as an action for the interacting CHS theory.
\foot{Notice that the resulting  classical nonlinear theory is a {\em metric-like} one  and does not require Vasiliev's unfolding procedure 
(though this alternative formulation should exist along the lines of \cite{Vasiliev:2009ck}).}

As we have found above,   
the bosonic CHS theory  has  vanishing Casimir energy,   
and  it may  also  be consistent at the quantum level  if in addition to the vanishing   of the  anomaly 
$a$-coefficient \cite{Giombi:2013yva,Tseytlin:2013jya}
the full conformal anomaly  also vanishes. This  theory  
may  thus be worth a detailed  exploration. 


\iffa
\begin{verbatim}
\end{verbatim}
\fi

\section*{Acknowledgments}
XB thanks T. Basile, N. Boulanger, E. Joung for useful exchanges on group-theoretic issues and is especially grateful to
O. Shaynkman for his patient introduction to \cite{Shaynkman:2004vu}.
AAT   is   grateful  to   S. Giombi,  R. Metsaev  and   M. Vasiliev  for  discussions.
The  work of AAT was supported by the ERC Advanced grant No.290456
and also by the STFC grant ST/J000353/1.
\appendix 

\section{Spectrum of   spin $s$  Laplacian  on $S^{d}$}
\label{A:Laplace}

Let us summarize the    result for the spectrum 
of the operator $
\OO_{s\, \perp}$ defined on symmetric
traceless transverse tensors of rank $s$ on $S^{d}$ 
\be
\OO_{s\, \perp}(M^{2}) \equiv (-D^{2}+M^{2})_{s\, \perp},\qquad\qquad 
\OO_{s\, \perp}\,\psi_{s}^{(n)} = \lambda_{n}\,\psi_{s}^{(n)}. \la{a1}
\ee
The eigenvalues and their degeneracy are given by  (see, e.g.,   \cite{Higuchi:1986wu,Camporesi:1994ga})
\begin{align}
\lambda_{n} &= (n+s)(n+s+d-1)-s+M^{2}, \qquad\ \ \ \ \  n=0,1,2,\dots, \la{a2} \\
\dd_{n} &= g_{s}\,\frac{(n+1)(n+2s+d-2)(2n+2s+d-1)(n+s+d-3)!}{(d-1)!(n+s+1)!}, \la{a3}\\
g_{s} &= \frac{(2s+d-3)(s+d-4)!}{(d-3)!s!}.\la{a4}
\end{align}
Here, $g_{s}$ is the number of components of the symmetric traceless transverse rank $s$ tensor
in $d$ dimensions.

\section{Expansions of curvature-squared  invariants}
\label{A:expan}

We used the following expressions  \cite{Barth:1983hb} for the quadratic terms in the background field expansion 
of the curvature-squared  terms appearing in the action of Weyl gravity
\be
\begin{split}
\delta^{2}\int d^{4}x\,\sqrt{g}\,R^{2} =& \int d^{4}x\,\sqrt{g}\Big[\frac{1}{2}\,R\,h_{\alpha\beta}\,D^{2}\,h^{\alpha\beta}
+R\,R_{\alpha\mu\beta\nu}\,h^{\alpha\beta}\,h^{\mu\nu}-\frac{1}{4}R^{2}\,h_{\alpha\beta}\,h^{\alpha\beta}
 \\
& +(R^{\alpha\beta}\,h_{\alpha\beta})^{2}+R\,R^{\alpha}_{\ \ \mu}\,h_{\alpha\beta}\,h^{\mu\beta}
+\dots\Big], 
\end{split}
\ee
\be
\begin{split}
& \delta^{2}\int d^{4}x\,\sqrt{g}\,R_{\mu\nu}R^{\mu\nu} = \int d^{4}x\,\sqrt{g}\Big[\frac{1}{4}\,h_{\alpha\beta}\,
D^{2}\,h^{\alpha\beta}
-\frac{1}{4}\,R_{\mu\nu}R^{\mu\nu}\,h_{\alpha\beta}\,h^{\alpha\beta} \\
& \qquad +R^{\alpha\mu\beta\nu}\,h_{\mu\nu}\,D^{2}\,h_{\alpha\beta}+\frac{1}{2}R^{\alpha\mu}\,R^{\beta\nu}\,h_{\alpha\beta}\,\,h_{\mu\nu}-\frac{1}{2}\,R^{\alpha\mu}\,R_{\mu\nu}\,h_{\alpha\beta}\,
h^{\beta\nu}+R_{\mu\nu}\,R^{\rho\mu\beta\nu}\,h_{\alpha\rho}\,h^{\alpha}_{\ \beta}  \\
&\qquad +R^{\alpha\mu\beta\nu}\,R_{\alpha\rho\beta\tau}\,h_{\mu\nu}\,h^{\rho\tau}
+\frac{1}{2}\,R^{\mu\nu}\,h_{\alpha\beta}\,D_{\mu}\,D_{\nu}\,h^{\alpha\beta}+\dots\Big],\no
\end{split}
\ee
where dots stand for terms with covariant derivatives of $R_{\mu\nu}$ or vanishing  due to the gauge conditions
\be
D^{\mu}\,h_{\mu\nu} = 0, \qquad\ \ \  h_{\mu}^{\mu}=0.
\ee
On conformally flat $d=4$ background we drop  also terms containing the Weyl tensor
\be
C_{\alpha\beta\gamma\delta} = R_{{\alpha\beta\gamma\delta}  } - \frac{1}{2}(g_{\alpha\gamma} R_{\delta\beta}-g_{\alpha\delta}R_{\gamma\beta}-g_{\beta\gamma}R_{\delta\alpha}+g_{\beta\delta}R_{\gamma\alpha})
-\frac{1}{6} R (g_{\alpha\delta}g_{\gamma\beta}-g_{\alpha\gamma}g_{\delta\beta}) \ .
\ee

\def \hO  {\hat \OO}
\section{Factorized  form of Weyl graviton operator on an   Einstein background}
\label{A:Einstein}
For generality, let us  also  record   the result  (see also  \ci{Deser:2012qg})  
of expansion of the 2nd-derivative action \rf{38} 
near a  generic Einstein  background $R_{\mu\nu}=\frac{1}{4}Rg_{\mu\nu}$ which is not
conformally-flat, {\em i.e.} may have a non-vanishing Weyl tensor (a simple example is $S^2 \times S^2$).
Then, the quadratic fluctuation action in \rf{310} is given by 
\begin{align}
&\mathscr L^{(2)}_{\phi\phi} = -\frac{1}{4}\,\phi_{\mu\nu}\,\phi^{\mu\nu}, \qquad\quad
\mathscr L^{(2)}_{h\phi} = \frac{1}{2}\phi^{\mu\nu}
D^{2}\, h_{\mu\nu}  -\frac{1}{12}R\,h^{\mu\nu}\phi_{\mu\nu}+R_{\mu\gamma\nu\delta}\,h^{\mu\nu}\phi^{\gamma\delta}
\ , \la{c1} \\
&\mathscr L^{(2)}_{hh}= -\frac{1}{144}R^{2}\,h_{\mu\nu}h^{\mu\nu}+\frac{1}{12}R\,R_{\mu\gamma\nu\delta}\,
h^{\mu\nu}h^{\gamma\delta}+\frac{1}{24}R\,h^{\mu\nu}D^{2}\,h_{\mu\nu}\ .\la{c2}
\end{align}
Suppressing tensor indices, we get
\ba
&\mathscr L^{(2)} = \frac{1}{2}\,h\,\hO \,\phi-\frac{1}{4}\phi\,\phi  + \frac{1}{24}R\,h\,\hO\,h\ , \la{c3}\\
&\hO = D^{2}+2\,\mathscr R-\frac{1}{6} R\ ,\qquad \qquad 
h\,\mathscr R\,h \equiv  h^{\mu\nu}\,R_{\mu\gamma\nu\delta}\,h^{\gamma\delta} \ .  \la{c4}
\end{align}
Integrating out $\phi$  we get  $\mathscr L^{(2)} ={1 \ov 4}  h\,\hO_{2}\,h, $   where   
\be
\hO_{2} = \big( 
\hO+\frac{1}{6}R\big)\,  \hO\ =  \big(-D^{2}-2\mathscr R\big)\big(-D^{2}-2\mathscr R+\frac{1}{6}R\big).
\la{c5}
\ee
In particular, for the conformally-flat Einstein space like $S^4$ or AdS$_4$ this  reduces to the operator  \rf{35}.

\label{sec:Metsaev3}

\iffa
\subsection{The case of $S^{4}$}
If we specialize on $S^{4}$, we have to set $R=12$ and $\mathscr R=-1$
because $(\mathscr R h)_{\mu\nu} = R_{\mu\gamma\nu\delta} h^{\gamma\delta} = (g_{\mu\nu}g_{\gamma\delta}-g_{\mu\gamma}g_{\nu\delta})\,h^{\gamma\delta} = -h_{\mu\nu}$.
Thus, we obtain 
\be
\mc O_{4} = \frac{1}{4}(-D^{2}+2)(-D^{2}+4),
\ee
in agreement with previous result (\ref{S4-fact}).
\fi


\section{On tensor zero modes  on $S^{3}$}
\label{A:zero2}

To illustrate the reason for the  truncation of the spectrum of $\OO_s$   operators in the CHS partition function  \rf{45} on 
$S^1 \times  S^3$, let us   consider explicitly the $s=2$ and $s=3$ cases.

In the   $s=2$ case we may  find the partition \rf{3116}  in the $h_{0\m}=0, \ h^\m_\m=0$   gauge 
where we are left with the spatial  traceless $h_{ij}$  tensor. 
We may split it as  $h_{ij}= h^{\rm TT}_{ij}  + D_{(i}V_{j)}$   where  $h^{\rm TT}_{ij}$ is  transverse traceless 
 and $V_i$ is  transverse,  $D^{i}V_{i}=0$. Let us consider $V_i$ to be the $n$-th eigenfunction of the 
Laplacian on $S^3$, {\em i.e.} $\mathbf{D}^2 V_i = -\l_{n} V_i$, where, according to \rf{a2}, we have 
$\l_n = (n+1)(n+3)-1$. We want to show that for the mode $n=0$, we have $D_{(i}V_{j)}=0$, {\em {\em i.e.}} $V_{i}$
is a Killing vector on $S^{3}$, and therefore this mode does not appear in the decomposition of $h_{ij}$.
Commuting the covariant derivatives on unit-radius $S^{3}$ one can show that 
\be
\label{D1}
(\mathbf{D}^{2}+\lambda_{n}-4)\,  D_{(i} V_{j)} =0\ ,\ \ \ \ \ \ \  D^{i}D_{(i}V_{j)} = (2-\lambda_{n})\,V_{j} \ . 
\ee
If $n=0$, then $\lambda_{n}=2$ and  thus  $D_{(i} V_{j)}$ is transverse. 
In this case, from \rf{a2} 
we see that $D_{(i} V_{j)}$ can be non-zero  only if the equation $(n'+2)(n'+4)-2 = -2$
admits a solution for some non-negative integer $n'$. Since this is impossible, we have proved 
the above statement.

The same mechanism works for spin 3. Now, we split the symmetric traceless field $\phi_{ijk}$ in 
TT part plus traceless components  as
\be
\label{D3}
\begin{split}
\phi_{ijk} &= \phi^{\rm TT}_{ijk}+ D_{(i}h^{\rm TT}_{jk)} + t_{ijk}, \\
t_{ijk} &= D_{(i}D_{j}V_{k)}-\frac{2}{5} g_{(ij}V_{k)}-\frac{1}{5} g_{(ij}\mathbf{D}^{2}\,V_{k)},
\end{split}
\ee
where 
$V_{i}$ is again transverse. 
We want to show that 
certain low modes of $h^{\rm TT}_{ij}$ and $V_{i}$ drop from this decomposition.
Let us begin with the first piece, $D_{(i}h^{\rm TT}_{jk)}$. Again, $h^{\rm TT}_{ij}$
can be assumed to be  the $n$-th eigenfunction of the 
Laplacian on $S^3$, {\em i.e.} $\mathbf{D}^2 h^{\rm TT}_{ij} = -\l_{n} h^{\rm TT}_{ij}$, where now 
$\l_n = (n+2)(n+4)-2$.  As in  (\ref{D1})  we get 
\be
\label{D4}
(\mathbf{D}^{2}+\lambda_{n}-6)\,D_{(k}h^{\rm TT}_{ij)}=0 \ .
\ee
 For the zero mode, $\lambda_{0}=6$ and $D_{(k}h^{\rm TT}_{ij)}=0$
follows from the positivity of the Laplacian and the fact that there are no constant 3-tensors. Thus, this mode 
drops out  from the splitting (\ref{D3}).

The analysis of the second term $t_{ijk}$ in (\ref{D3}) goes along similar lines.
We assume $V_{i}$ to be the $n$-th mode
of the Laplacian on $S^3$, factoriz $\mathbf{D}^2 V_i = -\l_{n} V_i$, with 
$\l_n = (n+1)(n+3)-1$. We must drop the zero mode of $V_{i}$ as  otherwise $t_{ijk}=0$,
due to the analysis of the spin 2 case.
To show that we must also drop the $n=1$ mode, we  use  the following relations, 
analogous to (\ref{D1}) and (\ref{D4}),
\be
\label{D6}
(\mathbf{D}^{2}+\lambda_{n}-10)\,t_{ijk}=0\ , \ \ \ \ \ \ \ \ \ 
D^{i}\,t_{ijk} = \frac{8}{5}(7-\lambda_{n})\,D_{(j}V_{k)} \ . 
\ee
This shows that   $t_{ijk}$ is transverse precisely at $n=1$ where $\lambda_{1} = 7$. In this case, taking into account 
that the spectrum of $-\mathbf{D}^{2}$ on 3-tensors is $(n'+3)(n'+5)-3$, we see that the operator 
$\mathbf{D}^{2}+\lambda_{1}-10 \equiv \mathbf{D}^{2}-3$ has no zero modes. Again, 
we conclude that $t_{ijk}=0$.


\def \rx {{\rm x}}

\section{Characters of $SO(d)$  representations}
\label{A:so-char}

Below we  present   the  basic  expression   for the character  of irreducible 
tensor representation  of 
$SO(d)$   with even $d=2r$  (for details, 
see, e.g., \cite{Dolan:2005wy}). 
The tensor representation is associated with a Young tableaux $\ell = (\ell_1, ..., \ell_r)$ with $r$ rows of 
length $\ell_{i}$, $i=1, \dots, r$. The character $\chi_{\ell}(\rx)$ is the following {\it  symmetric} function 
of $\rx=(x_{1}, \dots, x_{r})$ 
\be
\label{E1}
\chi_{\ell}(\rx) = \frac{\mathscr D_{+}(\rx)-\mathscr D_{-}(\rx)}{2\,\mathscr D(\rx)}\ ,
\ee
where 
\begin{align}
&\mathscr D_{\pm}(\rx) = \det(x_{i}^{k_{j}}\pm x_{i}^{-k_{j}}), \qquad \ \ \  k_{j}=\ell_{j}+r-j \ , \ \ \   \la{E2}\\
&\mathscr D(\rx) = \prod_{1\le i<j\le r}(x_{i}+x_{i}^{-1}-x_{j}-x_{j}^{-1})\ .  \la{E3}
\end{align}
One can show that   (\ref{E1}) leads to a polynomial in $x_{i}$ and $x_{i}^{-1}$ after an  algebraic
factorization. This follows from the general definition 
of the character as a trace.

 In the main text, we are interested in special  cases  where most of  components $x_{i}$ are  equal to 1.
This is a smooth unambiguous limit after factorization. 
If one starts directly  with  (\ref{E1}), one has to be careful to take a  limit   to 
avoid removable singularities, as we do in eq.(\ref{54}).

\section{Characters of relevant $\mso(d,2)$  representations}
\label{A:BGG}


Below    we  will   explain  the structure  of spaces  of  some representations 
(i.e. modules)
  of  the Lie 
algebra  $\mso(d,2)$ of the conformal group in $d$ dimensions  that  are relevant for the 
 computation of the  CFT partition functions in the main text.    By  kinematics of  vectorial AdS$_{d+1}$/CFT$_{d}$ , these representations  
 appear also 
in the description of  massless symmetric  higher  spin   $s\geqslant 1$ fields on AdS$_{d+1}$. 
We will  present the explicit expressions  for the associated characters.

The structure of all (generalized) Verma modules\,\foot{Verma modules are highest-weight modules, i.e. 
they are  generated by a highest weight vector.
Strictly speaking, in the present paper we consider \textit{generalized} Verma modules, i.e. freely generated by a (finite-dimensional) highest-weight \textit{space} but we will drop the term ``generalized'' since this distinction is rather technical.
}
 of $\mso(d,2)$ has been described in \cite{Shaynkman:2004vu}.
The fundamental results of \cite{Shaynkman:2004vu}
 are  applied   below  to  the particular $\mso(d,2)$-modules  we are interested in here.\footnote{More precisely, 
the general analysis presented in Subsection 4.4 of \cite{Shaynkman:2004vu} is particularized  here 
to the modules in the first series associated with the dominant integral weight $(\lambda_0,\lambda_1,\ldots,\lambda_d)=(s-1,s-1,0,\ldots,0)$ in the notation of \cite{Shaynkman:2004vu}.} 
While in the main text we consider only  the case of even $d$, here 
we include for completeness  also  the  case of odd $d$  which happens to be much simpler. 


\subsection{Characters of generalized Verma $\mso(d,2)$-modules}\label{Verma}
Let us first introduce some basic definitions and notation.
The commutation relations of the generators of $\mso(d,2)$ can be cast in the form 
\begin{eqnarray}
&&\left[\textsc{E}, \textsc{J}_i^{\pm}\right]=\pm  \textsc{J}_i^{\pm} ~~~,\ \ \ ~~~ \left[\textsc{J}_{ij},\textsc{J}^{\pm}_k\right]= 2i\delta_{k[j} \textsc{J}^{\pm}_{i]}\nonumber\\
&&\left[\textsc{J}^-_i, \textsc{J}^+_j\right]=2(i\textsc{J}_{ij}+\delta_{ij}\textsc{E})\label{comrel}\\
&&\left[\textsc{J}_{ij},\textsc{J}_{kl}\right]=i\delta_{jk}\textsc{J}_{il} \,+\, \mbox{antisymetrizations}\nonumber
\end{eqnarray}
where $i,j=1,2,\ldots,d$ and the ladder operators $\textsc{J}^{\pm}_i
$ shift the eigenvalues of $\textsc{E}$ by $\pm1$.

The maximal compact subalgebra of $\mso(d,2)$ is the direct sum
$\mso(2)\oplus\mso(d)=\mbox{span}\{\textsc{E},\textsc{J}_{ij}\}$.
The parabolic subalgebra
$
\mbox{span}\{\textsc{E},\textsc{J}^-_i,\textsc{J}_{jk}\}$ 
is isomorphic to the algebra of homotheties of the Euclidean space ${\mathbb R}^d$.
Consider the finite-dimensional irreducible module (to be denoted as  ${\cal Y}_{[\Delta,\ell]}$) of 
the latter algebra which is  characterized by the eigenvalue $\Delta$ of  $\textsc{E}$, annihilated by $\textsc{J}^-_i$ and carrying a finite-dimensional representation of $\mso(d)$
characterized by a weight  vector 
$
\ell\equiv(\ell_1,\ldots,\ell_{r-1},\ell_r)$.  Here $r$ denotes the rank of $\mso(d)$ 
(i.e. the integer part of $d/2$)
 and  the labels satisfy
$\ell_1\geqslant\ldots\geqslant \ell_{r-1}\geqslant |\ell_{r}|\geqslant 0$.
When $d$ is even, the last label $\ell_{r}$ can be positive or negative. Non-negative integer labels define a Young diagram where the length of the $i$-th row is $\ell_i$.\footnote{In order to remove the exception of negative labels and deal, in the sequel, only with non-negative labels 
$\ell_i\geqslant 0$ ($i=1,\ldots,r$), 
we introduce the non-standard notation $\ell_\pm\equiv (\ell_1,\ell_2,\ldots,\pm\ell_r)$
for the weights with nonvanishing last label when $d$ is even. 
With a slight abuse of notation, in this exceptional cases ${\cal Y}_{[\Delta,\ell]}$ 
will stand for the reducible module  ${\cal Y}_{[\Delta,\ell_+]}\oplus {\cal Y}_{[\Delta, \ell_-]}$. 
The reducibility is then interpreted as the fact that a chirality condition can be imposed, e.g. (anti)selfduality on the 
$\ell_{\frac{d}2}$ columns
of length $d/2$.
}

The Verma $\mso(d,2)$-module ${\cal V}_{[\Delta,\ell]}$ is the module freely generated from the module ${\cal Y}_{[\Delta,\ell]}$
by the action of the raising operators $\textsc{J}^+_i$.
Following standard usage in the literature, we will often make use of the CFT$_d$ language to describe the $\mso(d,2)$-modules
although, strictly speaking, this language is adapted to the decomposition with respect to the noncompact subalgebra $\mso(1,1)\oplus\mso(d-1,1)$ in which case the raising and lowering operators are the translation and the conformal boost generators. Modulo this slight abuse of terminology, the Verma module ${\cal V}\left(\Delta,\ell\right)$ 
can be interpreted as the module spanned by the primary conformal  field  of scaling dimension $\Delta$ 
 (or, equivalently, by ${\cal Y}_{[\Delta,\ell]}$ in the above notation)
  together with all its descendants. A descendant that is at the same time a primary  is called a singular module. 

The characters are generating functions of weight multiplicities via their power expansion in the variable $q$ corresponding to 
$\mso(2)$  generated by $\textsc{E}$
and of the variables 
$\rx=(x_{1}, \dots, x_{r})$
generated by  $\textsc{J}_{ij}$. 
We shall denote the character by the same label as the corresponding module. 
The character of  ${\cal Y}_{[\Delta,\ell]}$     can be written as 
 \be  {\cal Y}_{[\Delta,\ell]}(q,\rx)=q^\Delta\chi_{\ell}(\rx)  \ ,  \la{f2}\ee
where $\chi_{\ell}(x)$ is the usual character of the finite-dimensional $\mso(d)$-module labelled by $\ell$ 
and recalled    above in appendix~\ref{A:so-char}. 
The character of the Verma module ${\cal V}_{[\Delta,\ell]}$ 
 is entirely determined  by  the character of ${\cal Y}_{[\Delta,\ell]}$
and  by  the character of the Verma module for the trivial weight  ${\cal V}_{[0,0]}(q,\rx)\equiv P(q,\rx) $, i.e. 
\be
{\cal V}_{[\Delta,\ell]}(q,\rx)={\cal Y}_{[\Delta,\ell]}(q,\rx){\cal V}_{[0,0]}(q,\rx)=q^\Delta\chi_{\ell}(\rx)P(q,\rx)\ . 
\label{charVerma}
\ee 
The explicit expression  for $P(q,\rx)$  in even $d=2r$ is   (in odd  $d=2r+1$ there is an  extra factor of $(1-q)^{-1}$, 
see, e.g.,  \cite{Dolan:2005wy})
\be  
P(q,\rx) = \prod_{i=1}^r { 1 \ov  ( 1- q x_i) ( 1- {q x_i^{-1}})} \ . 
 \la{f4}\ee

\subsection{Odd dimension $d\geqslant 3$}

As usual,  one  may represent $\mso(d,2)$ as acting on  fields in $\mR^d$, i.e. 
interpret its  representations in CFT$_d$  language. 
The main representations  considered below will be the conserved  spin $s$ current one,   the 
conformal spin $s$ or shadow field one   and the conformal Killing tensor one. 

Let us denote by ${\cal D}_{[1-s;(s-1)]}$ the $\mso(d,2)$-module 
 associated with the  conformal Killing tensor fields  in flat  $d$-dimensional  space
which are totally symmetric tensors of rank $s-1$ and dimension $\Delta =1-s$
subject to 
the differential constraint: the traceless part of $\del_{(i_1} k_{i_2...i_s)}$  should vanish
    (cf. \cite{Eastwood:2002su}).
They  may be interpreted as  trivial gauge transformations of the conformal spin $s$ field $\p_s$ in \rf{2}
or  of the shadow field in CFT$_d$ 
(and thus zero modes of the corresponding  CHS ghost determinant). 
The   corresponding  finite-dimensional irreducible $\mso(d,2)$-module
 may be formally labelled also by 
 the 
rectangular Young diagram $(s-1,s-1,0,...,0)$  of $\mso(d+2)$ made of two rows of length $s-1$.

For odd dimension $d$, the resolution\,\footnote{Let us recall  few basic facts of homological algebra.
 A finite resolution of length $n\in\mathbb N$ of the module $V_0$ is an exact sequence of homomorphisms $d_i$
\be
0\stackrel{d_{n+2}}{\rightarrow} V_{n+1}\stackrel{d_{n+1}}{\rightarrow} V_n\stackrel{d_n}{\rightarrow}\cdots\stackrel{d_2}{\rightarrow} V_1\stackrel{d_1}{\rightarrow} V_{0}\stackrel{d_0}{\rightarrow}  0\,.\no
\label{resolution}
\ee 
A short exact sequence $0\rightarrow V_2\rightarrow V_1 \rightarrow V_0 \rightarrow 0$ is a resolution of length $1$
expressing  that $V_0=V_1/V_2$.
More generally, the exactness of the  above sequence,    
{\em i.e.} $\mbox{Ker}\, d_i=\mbox{Im}\, d_{i+1}\equiv D_i$ ($i=0,1,\ldots,n+1$), produces 
the following chain of short exact sequences
\be
0\rightarrow D_{i+1}\rightarrow V_{i+1}\rightarrow D_i\rightarrow 0\qquad(i=0,1,\ldots,n)\label{shortseqs}\no 
\ee
since $\mbox{Im}\, d_i=V_i/\mbox{Ker}\, d_i$.
Notice that the first and last short exact sequence are degenerate and simply express that $D_0=V_0$ and $D_n=V_{n+1}$. 
The other members in this chain 
determine recursively all modules $D_i$ in terms of the $V_j$ with $j>i$. In particular,
the module $V_0=D_0$ is resolved in the sense that $V_0=V_1/(V_2/\ldots(V_n/V_{n+1}))$.
} 
of ${\cal D}_{[1-s;(s-1)]}$ \`a la Bernstein-Gelfand-Gelfand is the following exact sequence of $\mso(d,2)$-modules (cf. appendix B of \cite{Shaynkman:2004vu}):
\be
\begin{split}
&0\rightarrow {\cal V}_{[s+d-1;(s-1)]}\rightarrow {\cal V}_{[s+d-2;(s)]}\rightarrow {\cal V}_{[d-2;(s,s)]}\rightarrow {\cal V}_{[d-3;(s,s,1)]}\\
&\rightarrow{\cal V}_{[d-4;(s,s,1^2)]}\rightarrow\cdots\rightarrow{\cal V}_{[\frac{d+1}2;(s,s,1^{\frac{d-5}2})]}\rightarrow{\cal V}_{[\frac{d-1}2;(s,s,1^{\frac{d-5}2})]}\rightarrow\cdots\rightarrow{\cal V}_{[4;(s,s,1^2)]}\\
&\rightarrow{\cal V}_{[3;(s,s,1)]}\rightarrow{\cal V}_{[2;(s,s)]}\rightarrow{\cal V}_{[2-s;(s)]}\rightarrow {\cal V}_{[1-s;(s-1)]}\rightarrow {\cal D}_{[1-s;(s-1)]}\rightarrow 0\,,
\label{BGGresolution}
\end{split}
\ee
where $(s,s,1^p)$ stands for a Young diagram with two rows of length $s$ and $p$ additional  rows of length 1. 
We also  use shorthand notation $(s)=(s,0,\dots,0)$ and $(s,s)=(s,s,0,\dots,0)$. 

Let us illustrate these homomorphisms by considering the first arrows in \eqref{BGGresolution}. The exactness implies that the map ${\cal V}_{[s+d-1;(s-1)]}\rightarrow {\cal V}_{[s+d-2;(s)]}$ is injective. In CFT$_d$ language, 
the module ${\cal V}_{[s+d-2;(s)]}$ is generated by a traceless symmetric tensor $j^{\mu_1\ldots\mu_s}$ of conformal dimension 
 $s+d-2$. 
Its
divergence $\partial_{\nu}j^{\nu\mu_1\ldots\mu_{s-1}}$ is a primary field of conformal dimension 
 $s+d-1$ and of rank $s-1$.
  The submodule generated by this primary field is indeed the image of 
	${\cal V}_{[s+d-1;(s-1)]}$ in ${\cal V}_{[s+d-2;(s)]}$.
Moreover, the exactness also implies that this submodule is mapped by
${\cal V}_{[s+d-2;(s)]}\rightarrow {\cal V}_{[d-2;(s,s)]}$ to zero. 
The module ${\cal V}_{[d-2;(s,s)]}$ is generated by tensors $k^{\mu_1\ldots\mu_s,\nu_1\ldots\nu_s}$ of dimension  $d-2$
with the symmetries of the spin-$s$ Weyl tensor. The $s$-th divergence $j^{\mu_1\ldots\mu_s}\equiv 
\partial_{\nu_1}\cdots\partial_{\nu_s} k^{\mu_1\ldots\mu_s,\nu_1\ldots\nu_s}$ is a primary field of  dimension  $s+d-2$ that can be interpreted as a trivial conserved current (an ``improvement''). The submodule generated by this primary field is the image of 
${\cal V}_{[s+d-2;(s)]}$ in ${\cal V}_{[d-2;(s,s)]}$, which
is isomorphic to the quotient module ${\cal D}_{[s+d-2;(s)]}={\cal V}_{[s+d-2;(s)]}/{\cal V}_{[s+d-1;(s-1)]}$ the representatives of which are conserved currents $j^{\mu_1\ldots\mu_s}$ since $\partial_{\nu}j^{\nu\mu_1\ldots\mu_{s-1}}\sim 0$ in 
${\cal D}_{[s+d-2;(s)]}$. In other words, any strictly conserved current is trivial.

The maximal submodule of a reducible Verma $\mso(d,2)$-module arises from singular modules for $d$ odd and the resolution 
\eqref{BGGresolution} of length 
$d$
provides the following recursive chain of irreducible modules \cite{Shaynkman:2004vu}: 
\begin{align}
{\cal D}_{[s+d-1;(s-1)]}&={\cal V}_{[s+d-1;(s-1)]}\ , \ \ \ \ \ \ \ \ 
{\cal D}_{[s+d-2;(s)]} ={\cal V}_{[s+d-2;(s)]}/{\cal D}_{[s+d-1;(s-1)]}\label{conslaw}\ , \\
{\cal D}_{[d-2;(s,s)]}&={\cal V}_{[2;(s,s)]}/{\cal D}_{[s+d-2;(s)]}\ , \ \ \ \ \ \ 
{\cal D}_{[d-3;(s,s,1)]}={\cal V}_{[3;(s,s,1)]}/{\cal D}_{[2;(s,s)]}\ , \ \ \ \   ... \\
{\cal D}_{[2;(s,s)]}&={\cal V}_{[2;(s,s)]}/{\cal D}_{[3;(s,s,1)]}\label{Weylc}\ , \ \ \ \ \ \ \ \ 
{\cal D}_{[2-s;(s)]}={\cal V}_{[2-s;(s)]}/{\cal D}_{[2;(s,s)]}\ ,  \\
{\cal D}_{[1-s;(s-1)]}&={\cal V}_{[1-s;(s-1)]}/{\cal D}_{[2-s;(s)]}\,.\label{confKillodd}
\end{align}
As we have seen, in CFT$_d$ language the module ${\cal D}_{[s+d-2;(s)]}$ is spanned by a conserved current of spin $s$ together with all its descendants and the quotient by ${\cal D}_{[s+d-1;(s-1)]}$ in \eqref{conslaw} is the translation of the conservation law. Similarly,  ${\cal D}_{[2;(s,s)]}$ 
describes the module generated by the  linearized Weyl-like   tensor 
  of a spin-$s$ shadow field while ${\cal D}_{[3;(s,s,1)]}$ in \eqref{Weylc} corresponds to the 
%
generalized 
  Bianchi identities. In turn, the module ${\cal D}_{[2-s;(s)]}$
describes the module for a \textit{pure gauge} shadow field since the quotient in \eqref{Weylc} means that the Weyl curvature is set to zero.
Finally, one recovers from \eqref{confKillodd} the initial identification of ${\cal D}_{[1-s;(s-1)]}$ as the module of conformal Killing tensor fields, since the quotient by ${\cal D}_{[2-s;(s)]}$ is the translation of the conformal Killing equation.

The chain \eqref{conslaw}-\eqref{confKillodd} allows to compute the characters of all irreducible modules, e.g., 
\be
\begin{split}
{\cal D}_{[s+d-2;(s)]}(q,\rx)=&\ \ {\cal V}_{[s+d-2;(s)]}- {\cal V}_{[s+d-1;(s-1)]}(q,\rx)\\
=&\ \ q^{s+d-2}(\,\chi_{(s)}(\rx)- q\,\chi_{(s-1)}(\rx)\,)\,P(q,\rx)\label{charconfcurr}\ , 
\end{split}\ee\be 
\label{charconfKillodd}
\begin{split}
{\cal D}_{[1-s;(s-1)]}(q,\rx) =&\ \  {\cal V}_{[1-s;(s-1)]}(q,\rx)+{\cal V}_{[s+d-1;(s-1)]}(q,\rx) \\
&-\left({\cal V}_{[2-s;(s)]}(q,\rx)+{\cal V}_{[s+d-2;(s)]}(q,\rx)\right)\\
&+\sum\limits_{p=0}^{\frac{d-5}2}(-1)^p({\cal V}_{[2+p;(s,s,1^p)]}(q,\rx)+{\cal V}_{[d-2-p;(s,s,1^p)]}(q,\rx))
\\
=&\Big[\,(q^{s+d-1}+q^{1-s})\chi_{(s-1)}(\rx) -(q^{s+d-2}+q^{2-s})\chi_{(s-1)}(\rx)\\
&\qquad\qquad+\sum\limits_{p=0}^{\frac{d-5}2}(-1)^p(q^{2+p}+q^{d-2-p})\chi_{(s,s,1^p)}(\rx)\, \Big]P(q,\rx)\ . 
\end{split}
\ee

Let ${\cal S}_{[2-s;(s)]}$ denote the $\mso(d,2)$-module generated by the shadow field of  dimension $2-s$   and spin $s$.
Notice that this module does not  appear in the list \eqref{conslaw}-\eqref{confKillodd}, although it is irreducible for $d$ odd.\footnote{The natural description of shadow fields
in this setting seems to be rather in terms of contragredient modules. In the computation of characters, one may ignore this subtlety.} 
Nevertheless, due to the previous identification of ${\cal D}_{[2-s;(s)]}$ with the module of gauge transformations for a spin-$s$ shadow field, 
one finds that the character of the shadow field itself is given by
\ba
{\cal S}_{[2-s;(s)]}(q,\rx)&={\cal V}_{[2-s;(s)]}(q,\rx)-{\cal D}_{[2-s;(s)]}(q,\rx)\la{ff1}\\
&={\cal V}_{[2-s;(s)]}(q,\rx)-{\cal V}_{[1-s;(s-1)]}(q,\rx)+{\cal D}_{[1-s;(s-1)]}(q,\rx)
\label{usefulidty}\\
&={\cal V}_{[s+d-1;(s-1)]}(q,\rx)-{\cal V}_{[s+d-2;(s)]}(q,\rx) \no \\
&\qquad +\sum\limits_{p=0}^{\frac{d-5}2}(-1)^p\Big[{\cal V}_{[2+p;(s,s,1^p)]}(q,\rx)+{\cal V}_{[d-2-p;(s,s,1^p)]}(q,\rx)\Big] \ , \la{f112}
\end{align}
where \rf{ff1} follows from the definition of the shadow field as a primary field modulo gauge symmetries, 
\rf{usefulidty}  comes from the  second  isomorphism in \eqref{confKillodd} and the value of the
 character \eqref{charconfKillodd}  has been used to obtain the third equality \rf{f112}.

\subsection{Even dimension $d\geqslant 4$}

For even dimension $d$, the situation is rather more intricate because  
the maximal submodule of a reducible Verma $\mso(d,2)$-module does not necessarily arises from singular modules only 
\cite{Shaynkman:2004vu}.
Moreover, the diagram of homomorphisms which is the analogue of \eqref{BGGresolution}
is not a mere line of arrows but a complicate diagram involving a rhombus in the middle and  nonstandard arrows (cf. appendix B of \cite{Shaynkman:2004vu}). 
For that reason, we will only provide the final result: the corresponding chain of irreducible modules. It has the same structure as
\eqref{conslaw}--\eqref{confKillodd} until its middle, but it differs in the lower half 
(cf. Subsection 4.4.2 of \cite{Shaynkman:2004vu}) 
\begin{align}
{\cal D}_{[s+d-1;(s-1)]}&={\cal V}_{[s+d-1;(s-1)]}\label{conslaweven}\ , \ \ \
{\cal D}_{[s+d-2;(s)]}={\cal V}_{[s+d-2;(s)]}/{\cal D}_{[s+d-1;(s-1)]}\ , \ \ \ \   ...  \\
{\cal D}_{[\frac{d}2+1;(s,s,1^{\frac{d}2-3})]}&={\cal V}_{[\frac{d}2+1;(s,s,1^{\frac{d}2-3})]}/{\cal D}_{[\frac{d}2+2;(s,s,1^{\frac{d}2-4})]}\ , \no \\
{\cal D}_{[\frac{d}2;(s,s,1^{\frac{d}2-2})_\pm]}&={\cal V}_{[\frac{d}2;(s,s,1^{\frac{d}2-2})_\pm]}/{\cal D}_{[\frac{d}2+1;(s,s,1^{\frac{d}2-3})]}\ , \no \\
{\cal D}_{[\frac{d}2;(s,s,1^{\frac{d}2-2})]}&={\cal U}_{[\frac{d}2;(s,s,1^{\frac{d}2-2})]}/{\cal V}^*_{[\frac{d}2+1;(s,s,1^{\frac{d}2-3})]}\ , \no \\
{\cal D}_{[\frac{d}2-1;(s,s,1^{\frac{d}2-3})]}
&={\cal V}_{[\frac{d}2-1;(s,s,1^{\frac{d}2-3})]}/{\cal U}_{[\frac{d}2;(s,s,1^{\frac{d}2-2})]}={\cal U}_{[\frac{d}2-1;(s,s,1^{\frac{d}2-3})]}/{\cal V}^*_{[\frac{d}2+2;(s,s,1^{\frac{d}2-4})]}\ , \ ...\no\\
{\cal D}_{[2;(s,s)]}&={\cal V}_{[2;(s,s)]}/{\cal U}_{[3;(s,s,1)]}={\cal U}_{[2;(s,s)]}/{\cal V}^*_{[s+d-2;(s)]}\ , \label{Weyleven}\\
{\cal D}_{[2-s;(s)]}&={\cal V}_{[2-s;(s)]}/{\cal U}_{[2;(s,s)]}
={\cal U}_{[2-s;(s)]}/{\cal V}^*_{[s+d-1;(s-1)]}\label{pgaugesheven}\ , \\
{\cal D}_{[1-s;(s-1)]}&={\cal V}_{[1-s;(s-1)]}/{\cal U}_{[2-s;(s)]}\  , \label{confKilleven}
\end{align}
where ${\cal V}^*$ stands for the contragredient\,\footnote{The contragredient module is defined in Subsection 4.3 of \cite{Shaynkman:2004vu}. Roughly speaking, it is the dual space where the role of raising and lowering operators are interchanged. In CFT language, the contragredient analogue of a primary field is a constant field since it is annihilated by the translations.
%
}
of the corresponding Verma module. The modules denoted by $\cal U$ correspond to reducible auxilliary modules.

In CFT$_d$ language, the module ${\cal D}_{[s+d-2;(s)]}$ and ${\cal D}_{[1-s;(s-1)]}$ keep their interpretations
 as  conserved current 
and conformal Killing tensor. However, the pure gauge shadow field corresponds now 
 to the auxilliary module ${\cal U}_{[2-s;(s)]}$  in \eqref{confKilleven}. The reason underlying this slight difference with $d$ odd case is that the module generated by a pure gauge field is reducible in even $d$ because there exist conformally-covariant gauge-fixing conditions.\footnote{The simplest example is $s=1$ 
where the following descendant of a pure gauge field $A_\mu=\partial_\mu\varepsilon$ is also a primary field:
$\Box^{\frac{d}2-1}\partial\cdot A=\Box^\frac{d}2\varepsilon$.}
Therefore, 
the irreducible module ${\cal D}_{[2-s;(s)]}$ can be interpreted as a pure gauge shadow field obeying 
a suitable gauge-fixing condition corresponding to the quotient by ${\cal V}^*_{[d+s-1;(s-1)]}$ in \eqref{pgaugesheven}.
Similarly, in odd dimension $d$ the off-shell Weyl-like  tensor of the spin-$s$ shadow field generates the irreducible module ${\cal D}_{[2;(s,s)]}$ but in even dimension $d$, the \textit{off-shell} Weyl-like  tensor is reducible 
since conformally-covariant equations of motion (corresponding to the submodule ${\cal V}^*_{[s+d-2;(s)]}$) in \eqref{Weyleven} can be imposed. Indeed, in even dimension $d$ the irreducible module ${\cal D}_{[2;(s,s)]}$ correspond to the \textit{on-shell} Weyl-like  tensor for a shadow field of spin $s$.

Given  the chain \eqref{conslaweven}--\eqref{confKilleven} of isomorphisms, one may again compute recusively the characters of all irreducible modules. 
The characters of the modules in the upper half of the chain 
are unchanged, e.g.,  \eqref{charconfcurr} holds, but the characters in the lower half
are slightly modified. For instance,
\begin{align}
{\cal D}_{[1-s;(s-1)]}(q,\rx) =& {\cal V}_{[1-s;(s-1)]}(q,\rx)+{\cal V}_{[s+d-1;(s-1)]}(q,\rx)\no  \\
 & \ \ -\left({\cal V}_{[2-s;(s)]}(q,\rx)+{\cal V}_{[s+d-2;(s)]}(q,\rx)\right)\no \\
&\qquad +\sum\limits_{p=0}^{\frac{d}2-3}(-1)^p\left({\cal V}_{[2+p;(s,s,1^p)]}(q,\rx)+{\cal V}_{[d-2-p;(s,s,1^p)]}(q,\rx)\right)\no \\
&\qquad +(-1)^{\frac{d}2}{\cal V}_{[\frac{d}2;(s,s,1^{\frac{d}2-2})]}(q,\rx) \label{charconfKilleven}\\
&=\Big[\,(q^{s+d-1}+q^{1-s})\chi_{(s-1)}(\rx) -(q^{s+d-2}+q^{2-s})\chi_{(s-1)}(\rx\no )\\
&\quad+\sum\limits_{p=0}^{\frac{d}2-3}(-1)^p(q^{2+p}+q^{d-2-p})\chi_{(s,s,1^p)}(\rx)+(-q)^{\frac{d}2}\chi_{(s,s,1^{\frac{d}2-2})}(\rx)\, \Big]P(q,\rx)
\no
\end{align}
Due to the identification of ${\cal U}_{[2-s;(s)]}$ with the module of gauge transformations for a spin-$s$ shadow field, 
one finds that the character of the module ${\cal S}_{[2-s;(s)]}$ generated by the shadow field itself  is given by
\be
{\cal S}_{[2-s;(s)]}(q,\rx) ={\cal V}_{[2-s;(s)]}(q,\rx)-{\cal U}_{[2-s;(s)]}(q,\rx)\,.\label{charshadoweven}
\ee

Notice that the important relation \eqref{usefulidty}  for    ${\cal S}_{[2-s;(s)]}(q,\rx)   $
(the counterpart of the prescription for $\Z_{-\,  s}$ proposed in the introduction)  holds also for $d$ even, as can be seen from \eqref{charshadoweven} and \eqref{confKilleven}.
The explicit expression for  this  character can be easily computed from \eqref{charVerma} and \eqref{charconfKilleven}.
By making use of the isomorphisms 
\eqref{Weyleven}--\eqref{confKilleven} together with the equalities \eqref{charconfcurr} and \eqref{charshadoweven},
one can prove  the  following remarkable identity (the counterpart of the relation in \rf{7})
\be
{\cal D}_{[2;(s,s)]}(q,\rx)={\cal S}_{[2-s;(s)]}(q,\rx)-{\cal D}_{[s+d-2;(s)]}(q,\rx)\ . 
\label{F19}
\ee
It is consistent with the expected isomorphism 
\be 
{\cal D}_{[2;(s,s)]}={\cal S}_{[2-s;(s)]}/{\cal D}_{[s+d-2;(s)]} \ , \la{F20}\ee
where the quotient by ${\cal D}_{[s+d-2;(s)]}$ is the translation of the imposition of the equations of motion
of order $d+2s-4$ on the spin-$s$ shadow field, or,  equivalently, of the conformal higher spin equations 
corresponding to \rf{2}  \cite{Fradkin:1985am,Segal:2002gd} (see also \cite{Vasiliev:2009ck,Bekaert:2012vt}). 
Another way to understand the appearance of
${\cal D}_{[s+d-2;(s)]}$ is that the natural source term for such equations is indeed a conformal current that we set to zero when we impose them.

The equations \eqref{charconfKilleven}, \eqref{charshadoweven} and \eqref{F19} allow us  to  find 
 the important character
of the  on-shell conformal field:
\begin{align}
&{\cal D}_{[2;(s,s)]}(q,\rx) =2\left({\cal V}_{[s+d-1;(s-1)]}(q,\rx)-{\cal V}_{[s+d-2;(s)]}(q,\rx)\right)
+(-1)^{\frac{d}2}{\cal V}_{[\frac{d}2;(s,s,1^{\frac{d}2-2})]}(q,\rx)\no 
\\
&\qquad\quad  +\sum\limits_{p=0}^{\frac{d}2-3}(-1)^p\left({\cal V}_{[2+p;(s,s,1^p)]}(q,\rx)+{\cal V}_{[d-2-p;(s,s,1^p)]}(q,\rx)\right)\no \\
&\qquad =\Big[\,2\,q^{s+d-2}\left(q\,\chi_{(s-1)}(\rx) -\chi_{(s-1)}(\rx)\right)\no  \\
&\quad\qquad+\sum\limits_{p=0}^{\frac{d}2-3}(-1)^p(q^{2+p}+q^{d-2-p})\chi_{(s,s,1^p)}(\rx)+(-q)^{\frac{d}2}\chi_{(s,s,1^{\frac{d}2-2})}(\rx)\, \Big]P(q,\rx)
\ . \label{F32}
\end{align}
Finally, let us note  also that  there is  the following relation between the relevant characters of 
$\mathfrak{so}(d,2)$  and $\mathfrak{so}(d+2)$ for the finite-dimensional representations
\be 
{\cal D}_{[1-s;(s-1)]}(q,x_1,\ldots,x_\frac{d}2)=\chi_{(s-1,s-1)}(q,x_1,\ldots,x_\frac{d}2) \ , \la{F22}
\ee
where  $\chi_{(s-1,s-1)}(y_1,y_2,\ldots,y_\frac{d}2,y_\frac{d+2}2)$ is the 
character of 2-row  $(s-1,s-1) \equiv (s-1,s-1,0,...,0)$  representation of  $\mathfrak{so}(d+2)$ 
defined in appendix \ref{A:so-char}.

\newpage

\iffa
\section{Temperature inversion properties}

The energy is computed from the one particle partition function $\Z(q)$ by means of 
\be
E(\beta) = -\frac{\partial}{\partial\beta}\,\log Z = -\frac{\partial}{\partial\beta}\,\sum_{m=1}^{\infty}\frac{1}{m}
\Z(q^{m}).
\ee
Hence, a term $q^{a}$ in $\Z(q)$ contributes $a/(e^{\beta\,a}-1)$ to the energy. For the CHS fields in $d=4$, we have
\be
\Z_{s}(q) =\sum_{n=0}^{\infty}\binom{n+3}{3}\bigg[2s^{2}\,q^{n+s+3}-2(s+1)^{2}\,q^{n+s+2}+2\,(2s+1)\,q^{n+2}\bigg],
\ee
and the associated energy is 
\be
E_{s}(\beta) =\sum_{n=0}^{\infty}\binom{n+3}{3}\bigg[2s^{2}\,\frac{n+s+3}{e^{\beta(n+s+3)}-1}
-2(s+1)^{2}\,\frac{n+s+2}{e^{\beta(n+s+2)}-1}
+2\,(2s+1)\,\frac{n+2}{e^{\beta(n+2)}-1}\bigg],
\ee
This can be rewritten as 
\be
\label{En}
\begin{split}
E_{s}(\beta) &= \frac{s(s+1)}{3}\,\sum_{n=1}^{\infty}\frac{n\,(3\,n^{2}-s^{2}-s-1)}{e^{\beta\,n}-1}+e_{s}(\beta),\\
e_{s}(\beta) &= -\frac{1}{3}\sum_{p=1}^{s+1} \frac{p}{e^{\beta\,p}-1}\,\bigg[
(2s+1)\,p^{3}-3s(s+1)\, p^{2}-(2s+1)\,p+s(s+1)(s^{2}+s+1)
\bigg] ,
\end{split}
\ee
where the second term $e_{s}(\beta)$ is a finite sum vanishing for $s=1$. It is convenient to split the infinite sum 
as follows
\be
\label{En2}
E_{s}(\beta) = \frac{s(s+1)}{3}\,\left[3\,f_{2}(\beta)-(s^{2}+s+1)\,f_{1}(\beta)\right]+e_{s}(\beta),
\ee
where
\be
f_{p}(\beta) = \sum_{n=1}^{\infty}\frac{n^{2p-1}}{e^{\beta\,p}-1}.
\ee
The functions $f_{p}(\beta)$ are non trivial at small $\beta$, but have a trivial expansion at large $\beta$:
\be
f_{p}(\beta) = e^{-\beta}+(2^{2p-1}+1)\,e^{-2\,\beta}+(3^{2p-1}+1)\,e^{-3\,\beta}+(2^{2p-1}+4^{2p-1}+1)\,e^{-4\,\beta}+\cdots\,.
\ee
In particular, they are exponentially suppressed in this regime. Instead, the small $\beta$ behaviour is governed by a temperature inversion relation thanks to the Ramanujan identity \cite{Dowker:2002ax}
\be
\label{Rama}
\begin{split}
f_{p}(\beta) &= \left(-\frac{4\,\pi^{2}}{\beta^{2}}\right)^{p}\,f_{p}\left(\frac{4\,\pi^{2}}{\beta}\right) 
+\bigg[1-\left(-\frac{4\,\pi^{2}}{\beta^{2}}\right)^{p}\bigg]\,\frac{B_{2p}}{4p}.
\end{split}
\ee
Remarkably, we can apply these identities thanks to the non trivial fact that the powers of $n$ in the numerator of the summand in the infinite summation are odd. 
Applying (\ref{Rama}) to our case, we find the small $\beta$ expansion of the first part of (\ref{En}),
\be
\label{part1}
\begin{split}
& \frac{s(s+1)}{3}\,\left[3\,f_{2}(\beta)- (s^{2}+s+1)\,f_{1}(\beta)\right] = \frac{\pi^{4}\,s(s+1)}{15\,\beta^{4}}
-\frac{\pi^{2}\,s(s+1)(s^{2}+s+1)}{18\,\beta^{2}}\\
& \qquad -\frac{1}{720}s(s+1)(10s^{2}+10s+13)+\mc O(e^{-1/\beta})
\end{split}
\ee
The small $\beta$ expansion of $e_{s}(\beta)$ is easily found by expanding the summand and computing the 
finite sum over $p$. This gives
\be
\label{part2}
\begin{split}
e_{s}(\beta) &= \frac{(s-1) s (s+1) \left(2 s^2+3 s+2\right)}{12 \beta }-\frac{1}{120}
   (s-1) s (s+1) (s+2) \left(3 s^2+3 s+2\right)\\
   & +\frac{\beta  (s-1) s^2
   (s+1)^2 (s+2) (2 s+1)}{1080}\\
   & -\frac{\beta ^3 \left((s-1) s^2 (s+1)^2
   (s+2) (2 s+1) \left(9 s^2+9 s-26\right)\right)}{1814400}+\mc O(\beta^{5}).
\end{split}
\ee
Notice that all the corrections involve odd powers of $\beta$. Summing (\ref{part1}) and (\ref{part2}), we obtain 
\be
\begin{split}
E_{s}(\beta) &= \frac{\pi ^4 s (s+1)}{15 \beta ^4}-\frac{\pi ^2 s (s+1)
   \left(s^2+s+1\right)}{18 \beta ^2}+\frac{(s-1) s (s+1) \left(2 s^2+3
   s+2\right)}{12 \beta }\\
   & -\frac{1}{720} s (s+1) \left(18 s^4+36 s^3+4
   s^2-14 s-11\right)\\
   & +\frac{\beta  (s-1) s^2 (s+1)^2 (s+2) (2
   s+1)}{1080}\\
   & -\frac{\beta ^3 \left((s-1) s^2 (s+1)^2 (s+2) (2 s+1)
   \left(9 s^2+9 s-26\right)\right)}{1814400}+\mc O(\beta^{5})+\mc O(e^{-1/\beta}).
\end{split}
\ee
The constant term is $-E_{c,s}$ and cancels in the sum $E_{s, \rm tot} = E_{s}(\beta)+E_{c,s}$ for all $s$, as already noticed in \cite{Gibbons:2006ij} for the cases $s=0,1$. The $1/\beta$ term is zero for $s=0,1$, but not for $s\ge 2$.
The inversion property of $E_{s}(\beta)$  for generic $s$ is complicated by the $e_{s}(\beta)$ term that is nevertheless
an elementary function. Apart from it, temperature inversion is completely captured by the relations (\ref{Rama})
for the non trivial part.

\fi

\bibliography{CHS-Biblio}{}
\bibliographystyle{JHEP}

\end{document}